%% file: Article.tex
\documentclass[11.5pt]{article}

\usepackage[utf8]{inputenc}
\usepackage[margin=1.0in]{geometry}

\usepackage{upgreek}
\usepackage{graphicx}
\usepackage{float}
\graphicspath{ {Images/} }

\usepackage{float}
\usepackage{cite}
\usepackage{braket}
\usepackage{amsmath}
\usepackage{gensymb}
\usepackage{multicol}
\usepackage{authblk}
\usepackage{mathtools}
\usepackage{multirow}
\usepackage{amsfonts}

\title{Explicit models of motions to analyze NMR relaxation data in proteins}
\author[1]{Nicolas Bolik-Coulon}
\author[1]{Fabien Ferrage}

\affil[1]{Laboratoire des Biomolécules, LBM, Département de Chimie, École Normale Supérieure, PSL University, Sorbonne Université, CNRS, 75005 Paris, France}

\newcommand{\beginsupplement}{%
        \setcounter{table}{0}
        \renewcommand{\thetable}{S\arabic{table}}%
        \setcounter{figure}{0}
        \renewcommand{\thefigure}{S\arabic{figure}}%
        \setcounter{equation}{0}
        \renewcommand{\theequation}{S\arabic{equation}}%
     }

\date{}
\begin{document}
\maketitle

\begin{abstract}
Nuclear Magnetic Resonance (NMR) is a tool of choice to characterize molecular motions. In biological macromolecules, pico- to nano-second motions, in particular, can be probed by nuclear spin relaxation rates which depend on the time fluctuations of the orientations of spin interaction frames. For the past 40 years, relaxation rates have been successfully analyzed using the Model Free (MF) approach which makes no assumption on the nature of motions and reports on the effective amplitude and time-scale of the motions. However, obtaining a mechanistic picture of motions from this type of analysis is difficult at best, unless complemented with molecular dynamics (MD) simulations. In spite of their limited accuracy, such simulations can be used to obtain the information necessary to build explicit models of motions designed to analyze NMR relaxation data. Here, we present how to build such models, suited in particular to describe motions of methyl-bearing protein side-chains and compare them with the MF approach. We show on synthetic data that explicit models of motions are more robust in the presence of rotamer jumps which dominate the relaxation in methyl groups of protein side-chains. We expect this work to motivate the use of explicit models of motion to analyze MD and NMR data.
\end{abstract}

\input{Section/1_Introduction}

\input{Section/2_BWR}

\input{Section/3_Models}

\input{Section/4_CorrelatedMotions}

\input{Section/5_CSAjumpRelaxation}

\input{Section/6_Discussion}

\input{Section/7_Conclusion}

\bibliographystyle{ieeetr}
\bibliography{Bibliography}

\newpage

\input{Section/SI}

\end{document}

%% file: Section/1_Introduction.tex
\section{Introduction}
Proteins are not rigid molecules and are better represented by ensembles of accessible conformations.\cite{Frauenfelder_Science_1991} The distributions of conformations and timescales of transitions between them report on the thermodynamics and kinetics of biomolecular processes. Nuclear Magnetic Resonance (NMR) is particularly suited for the characterization of dynamic properties at the atomic level with a large toolset to probe timescales ranging from picoseconds to seconds and more \cite{Charlier_ChemSocRev_2016,Alderson_Cell_2021}. Fast local motions of protein backbone and side-chains, on the pico- and nanosecond range, are of special interest as they contribute to allosteric pathways \cite{Choy_MolCell_2017}, catalytic reaction \cite{Otten_Science_2020}, the malleability of binding sites to accomodate different ligands or, in the case of side-chains, act as an entropy reservoir \cite{DuBay_AccChemRes_2015,Kasinath_ANIE_2015}.\\
A large variety of NMR experiments has been developped to measure the rates at which ensembles of nuclear spin return to their equilibrium after a perturbation \cite{Palmer_ChemRev_2004}, a process called nuclear spin relaxation. Relaxation is driven by intra- and inter-atomic interactions which strengths are affected by - or correlated to - the fluctuations of bond vectors orientations \cite{Nicholas_PNMRS_2010}. Thus, relating relaxation rates to the time-dependent orientations of the interaction frames allows to characterize the dynamics of bond vectors. This can be quantified in the frame of the Bloch-Wangsness-Redfield (BWR) relaxation theory which, in addition to quantum mechanic terms related to spin interactions, uses spherical harmonics to describe the orientation of the interaction frames in a fixed laboratory frame \cite{Wangsness_PhysRev_1953,Redfield_IBM_1957,Redfield_AdvMAgnOptReson_1965}. \\
Relaxation properties depend on the correlation function for the orientation of the principal axis frames of interactions contributing to relaxation. More specifically, relaxation rates are expressed as linear combinations of its Fourier transform, called the spectral density function, and evaluated at the eigenfrequencies of the spin system. In order to obtain a quantitative description of the motions from the analysis of NMR relaxation rates, an analytical form of the correlation function has to be written with parameters describing the amplitudes and timescales of the motions. This requires \textit{a priori} knowledge on the type of motions to build a model of correlation function, and can lead to complex analytical expressions. In addition, when analyzing a small set of experimental relaxation rates, several physically different models may describe the experiments equally well. \\
These limitations have been overcome by the Model-Free (MF) approach which aims at reproducing the correlation function for internal motions using a decaying exponential characterized by an effective correlation-time and a limit at infinite time called generalized order parameter \cite{Lipari_JACS_1982_1}. This model has later been complexified to the Extended Model-Free (EMF) approach \cite{Clore_JACS_1990} to include two time-scales and accompagnying order parameters. Both have lead to a large number of succesfull studies of protein dynamics by NMR \cite{Lipari_JACS_1982_2,Nicholson_biochemistry_1992,Farrow_Biochemistry_1994,Mandel_JMB_1995, Fushman_JACS_1999,Jarymowycz_CR_2006}. However, MF or EMF approaches are uninformative on the nature of motions of protein backbone and side-chains, and relaxation analysis has to be complemented with Molecular Dynamics (MD) simulations \cite{LindorffLarsen_Nature_2005,Robustelli_JACS_2012,Salvi_JPCL_2016,Cousin_JACS_2018,Smith_anie_2019, Kummerer_JCTP_2020}. \\
In MF type of analysis, the global tumbling is often supposed to be isotropic, which is the frame in which it was originially presented \cite{Lipari_JACS_1982_1}. An approximated correlation function for overall diffusion can be used when the diffusion tensor is anisotropic \cite{Lipari_JACS_1982_1,Tjandra_JACS_1995}. This limitation has also raised a number of questions regarding the relevance of the analysis of relaxation data with MF-type of correlation functions in the case of couplings between global (\textit{i.e} overall diffusion) and local motions \cite{Tugarinov_JACS_2001,Meirovitch_JPCB_2003}. and has lead to the development of the Slowly Relaxing Local Structure(SRLS) approach for the analysis of NMR relaxation data \cite{Polimeno_JPC_1995,Liang_JPCB_1999}. In addition, and as suggested by Freed and coworkers, the assumptions of simple local geometry and motions with axial symmetry have to be invoked in the frame of  the MF and EMF approaches \cite{Meirovitch_JPCA_2006,Meirovitch_JPCB_2006,Meirovitch_JPCB_2007}. This aspect is particularly critical as MF order parameters have been shown to be related to entropy, both for protein backbone \cite{Yang_JMB_1996} and side-chains \cite{Li_JACS_2009}, and a mischaracterization of the motions will inevitably lead to a mischaracterization of the thermodynamics \cite{Hoffmann_JPCB_2022}. \\ 
Although perfectible, modern MD simulations inform us quantitatively on the nature of molecular motions. It is now possible to obtain the \textit{a priori} knowledge required to develop explicit models of motions, building on analytical models introduced between 1960 and 1980 \cite{Wallach_JCP_1967,Witterbort_JCP_1978,Wang_JCP_1980}. Thus, the complexity of the expressions of correlation functions is now outweight by the wealth of information that can potentially be obtained by using such models. Here, we review three models (the rotation on a cone, the rotamer jump and the wobbling in a cone) and detail how to write complex correlation functions adapted to protein backbone and side-chain dynamics. We discuss cases of correlated motions and cases where the strengths of the nuclear spin interactions include time-dependent amplitudes. We aim at presenting a coherent framework to write correlation functions suitable for a wide range of combinations of motions. We use synthetic data to compare explicit models of motions and the MF correlation function, in particular in the presence of asymmetric internal and overall motions. This work lays the theoretical fundation for the analysis fo experimental NMR relaxation rates and molecular dynamics simulations, which we present elsewhere \cite{BolikCoulon_ArXiv_2022}.

%% file: Section/2_BWR.tex
\section{Correlation function in the BWR theory}
The description of the Bloch-Wangsness-Redfield (BWR) relaxation theory is beyond the scope of this article and can be found  elsewhere \cite{Goldman_JMR_1984,Cavanagh2007,Nicholas_PNMRS_2010}. We will only write the steps essential to reveal the correlation function in the relaxation superoperator. The evolution of the density operator $\hat{\sigma}$ is given by the Liouville-von Neumann equation:
\begin{equation}
\frac{\mathrm{d} \hat{\sigma}}{\mathrm{d}t}  = -i \left[ \hat{\mathcal{H}}(t),  \hat{\sigma}(t) \right],
\label{eq:Liouville}
\end{equation}
where the Hamiltonian $\hat{\mathcal{H}}(t) = \hat{\mathcal{H}}_0 + \hat{\mathcal{H}}_1(t)$ is expressed as the sum of stationnary $\hat{\mathcal{H}}_0$ and fluctuating $\hat{\mathcal{H}}_1(t)$ Hamiltonians. This equation is transformed in the interaction frame of the Hamiltonian $\hat{\mathcal{H}}_0$ (denoted by a tilde) to yield, after considering the ensemble average (denoted by the overbar) and hypotheses introduced by Bloch, Wangsness and Redfield:
\begin{equation}
	\frac{ \overline{\mathrm{d} \tilde{\hat{\sigma}}(t)}}{\mathrm{d} t}=- \int \limits_0^{\infty} \overline{ \left[ \tilde{\hat{\mathcal{H}}}_1(t),[\tilde{\hat{\mathcal{H}}}_1(t+\tau),\tilde{\hat{\sigma}}(t) - \tilde{\hat{\sigma}}_{eq}] \right]}\mathrm{d} \tau,
    \label{eq:MasterEquation}
\end{equation}
where $\tilde{\hat{\sigma}}_{eq}$ is the equilibrium density operator. The BWR relaxation theory is a semi-classical formalism based on the separation of the quantum spin part and the geometrical part when writting the Hamiltonian for the interactions contributing to a relaxation mechanism:
\begin{equation}
	\hat{\mathcal{H}}_1(t)=\sum_{i}\zeta_i \sum^{2} \limits_{q=-2}(-1)^q V^i_{2,-q}(t) \hat{T}^i_{2,q}, 
    \label{eq:Hpertube}
\end{equation}
where $\zeta_i$ is the amplitude of the interaction $i$, $V^i_{2,-q}(t)$ is related to a rank-2 Wigner matrix and $\hat{T}^i_{2,q}$ is a rank-2 tensor spin operator with coherence order $q$ and which can further be decomposed as a sum of eigenoperators ${\hat{A}_{2,q,p}^i}$ of the super-operator $[\hat{\mathcal{H}}_0, \cdot]$ with eigenvalues $\omega_{2,q,p}^{(i)}$. After subsituting Eq.\,\ref{eq:Hpertube} in Eq.\,\ref{eq:MasterEquation}, we obtain:
\begin{equation}
	\begin{aligned}
	\frac{\overline{\mathrm{d}  \tilde{\hat{\sigma}}(t)}}{\mathrm{d} t} = &
              - \sum \limits_{i,j} \zeta_i \zeta_j \sum^{2} \limits_{q,q'=-2}  \sum_{p, p'} (-1)^{q+q'}   e^{i(\omega_{2,q,p}^{(i)} - \omega_{2,q',p'}^{(j)}) t}   \left[\hat{A}_{2,q,p}^{i},[\hat{A}_{2,q',p'}^{j,\dag},\tilde{\hat{\sigma}}(t)] \right] \times \\
		&  \int \limits_0^{\infty} \langle V^i_{2,-q}(t)V_{2,-q'}^{j,\ast}(t+\tau) \rangle e^{-i\omega_{2,q',p'}^{(j)}\tau} \mathrm{d} \tau,
	\end{aligned}
\end{equation}
where $\langle \cdot \cdot \cdot \rangle$ denotes an ensemble average and $\dag$ and $*$ stand for the conjugate transpose and complex conjugate, respectively. The correlation function $C_{i,j}$ between interactions $i$ and $j$ is now defined as:
\begin{equation}
\langle V^i_{2,-q}(0)V_{2,-q'}^{j,\ast}(t) \rangle = \delta_{q,q'} C_{i,j} (t).
\end{equation}
It can be written using the Wigner matrix notations:
\begin{equation}
C_{i,j}(t) = \langle \mathcal{D}_{q,0}^{(2)*} (\Omega_{L,i}, 0) \mathcal{D}_{q,0}^{(2)} (\Omega_{L,j}, t) \rangle,
\label{eq:CorrelationFunctionBWR}
\end{equation}
where $\Omega_{L,i} = \{\alpha_{L,i}, \beta_{L,i}, \gamma_{L,i}\}$ is the set of three Euler angles orienting the interaction frame in the laboratory frame. 

\paragraph*{Useful properties of Wigner matrices.}
Before going any further, we will review some mathematical properties of Wigner rotation matrices that will be used when calculating correlation functions. \\
Rank-2 spherical harmonics $\mathrm{Y}_{2m}$ are directly linked to Wigner matrices $\mathcal{D}_{m0}^{(2)}$ as:
\begin{equation}
\mathcal{D}_{m0}^{(2)} (\alpha, \beta, \gamma) = \sqrt{\frac{4\pi}{5}} \mathrm{Y}_{2m}^* (\beta, \alpha),
	\label{eq:WignerSphericalH}
\end{equation}
with:
\begin{equation}
\mathcal{D}_{mn}^{(2)} (\alpha, \beta, \gamma) = e^{-i m \alpha} d_{mn}^{(2)} (\beta) e^{-i n \gamma},
	\label{eq:expWigner}
\end{equation}
where $d_{mn}$ is called the reduced Wigner matrix. From this definition (and the one of the reduced Wigner matrices), it follows:
\begin{equation}
\mathcal{D}_{mn}^{(2)} (\Omega) = (-1)^{-m-n} \mathcal{D}_{-m-n}^{(2)*} (\Omega).
\label{eq:PropConjWigner}
\end{equation}
The addition theorem relates rank-2 spherical harmonics (or Wigner matrices) to the second-order Legendre polynomial $\mathcal{P}_2 (x) = \frac{3 x^2 -1}{2}$:
\begin{equation}
\mathcal{P}_2 (\vec{x} \cdot \vec{y}) = \sum_{m=-2}^2 \mathcal{D}_{m0}^{(2)*} (\vec{y}) \mathcal{D}_{m0}^{(2)} (\vec{x}).
\label{eq:SumProdWignerP2}
\end{equation}
Wigner matrices can be decomposed into succesive rotations:
\begin{equation}
\mathcal{D}_{mn}^{(2)} (\Omega_{A, B}) = \sum_{x_1=-2}^2 \sum_{x_2=-2}^2 ... \sum_{x_k=-2}^2 \mathcal{D}_{mx_1}^{(2)} (\Omega_{A, X_1})   \mathcal{D}_{x_1 x_2}^{(2)} (\Omega_{X_1, X_2}) ... \mathcal{D}_{x_{k} n}^{(2)} (\Omega_{X_{k}, B}),
	\label{eq:PropDecomposition}
\end{equation}
where $\Omega_{X_i, X_j}$ is the Euler angle for transformation between frame $X_i$ and frame $X_j$.\\
Wigner matrices can be normalized:
\begin{equation}
\int d \Omega \mathcal{D}_{mn}^{L*} (\Omega) \mathcal{D}_{m'n'}^{L'} (\Omega) = \frac{8 \pi^2}{2L + 1} \delta_{LL'} \delta_{mm'} \delta_{nn'},
\label{eq:PropOrtho}
\end{equation}
where $\delta$ is the Kronecker function. \\
Finally, we have the following property:
\begin{equation}
\sum_{m=-2}^2 \mathcal{D}_{mn}^{(2)*} (\alpha, \beta, 0)\mathcal{D}_{mn'}^{(2)} (\alpha, \beta, 0) = \delta_{nn'}  .
\label{eq:SumProdWigner}
\end{equation}
It is worthwhile to  specify the convention  for Euler  angle used throughout this paper. If we write an Euler angle set $\Omega=\{\alpha, \beta, \gamma\}$, for tranformation from a frame $\{O,x, y, z\}$ to $\{O', x', y', z'\}$ (the two frame origins can be different, which only yields a translation without affecting the rotations), $\alpha$ is the  rotation angle around the $Oz$ axis, $\beta$ is  the rotation angle about the new $Ox$ axis (axis rotated by an angle $\alpha$ around the $Oz$ axis), and $\gamma$ is the rotation about the  new $Oz$ axis (after applying the two  previous rotations, only the last one affecting the z-axis). \\

%% file: Section/3_Models.tex
\section{Explicit models of correlation function}

\subsection{Approach to calculate correlation functions}
The approach to calculate the ensemble average in Eq.\,\ref{eq:CorrelationFunctionBWR} is detailed in Ref.\,\cite{Luginbuhl_PNMRS_2002}. If we write:
\begin{equation}
C (t) = \langle A^* (\Omega(0)) B (\Omega (t)) \rangle,
\end{equation}
with $A$ and $B$ two Wigner matrices with potentially different indices, the ensemble average expands into:
\begin{equation}
C (t) = \int d \Omega_0 \int d \Omega P(\Omega_0) P(\Omega, t | \Omega_0, 0) A^* (\Omega_0) B (\Omega),
\label{eq:PropIntegrationC}
\end{equation}
where $P(\Omega)$ is the probability that the Euler angles for frame transformation is $\Omega$, and $P(\Omega, t | \Omega_0, 0)$ is the conditional probability that the Euler angle for frame transformation is $\Omega$ at time $t$ when it was $\Omega_0$ at time $t=0$. In the following, we will calculate the ensemble averages using a Master equation for the description of relaxation-inducing processes:
\begin{equation}
\frac{\partial}{\partial t} P(\Omega, t) = -K P(\Omega, t),
\label{eq:CorrFuncApproachOp}
\end{equation}
which is a Fokker-Planck diffusion equation and where $P(\Omega, t)$ is the probability of finding the Euler angles $\Omega$ for frame transformation, at time $t$, and $K$ is a model-dependent operator describing the motions of interest. The initial step is to diagonalize the operator $K$, with eigenvalues $E_n$ and eigenvectors $\psi_n$:
\begin{equation}
K \psi_n = E_n \psi_n.
\label{eq:CorrFuncApproachEigEq}
\end{equation}
The conditional probability is then:
\begin{equation}
P(\Omega, t | \Omega_0, 0) = \sum_n \psi_n^* (\Omega_0) \psi_n (\Omega) e^{-E_n t},
\label{eq:PropCondProb}
\end{equation}
and:
\begin{equation}
P(\Omega_0) = \lim_{t \rightarrow \infty} P(\Omega, t | \Omega_0, 0).
\label{eq:PropProbInit}
\end{equation}
$P(\Omega_0)$ can sometimes be evaluated by simple probabilistic and geometric considerations. These probabilities are then inserted in Eq.\,\ref{eq:PropIntegrationC} to calculate the correlation function.

\subsection{Global tumbling}
\label{section:GlobalTumblingCorrFunc}
In this section, we will only consider the rotational diffusion of the protein: all interactions are fixed in the molecular frame. The correlation function describes the motions of the principal axis frames of interactions in the laboratory frame, which is equivalent to the motions of the protein in the laboratory frame. We follow the thorough treatment of this situation presented by Werbelow and Grant \cite{Werbelow_AdvMAgnOptReson_1977}, but other approaches can be found elsewhere \cite{Favro_PhysRev_1960, Woessner_JCP_1962, Stleele_JCP_1963, Huntress_JCP_1968}, \\
The correlation function can be decomposed using Eq.\,\ref{eq:PropDecomposition}:
\begin{equation}
	\begin{aligned}
C_{i,j} (t) = & \langle  \mathcal{D}_{q0}^{(2)*} (\Omega_{L,i},0) \mathcal{D}_{q0}^{(2)} (\Omega_{L,j},t) \rangle \\
		=& \sum_{a,a'=-2}^2  \langle  \mathcal{D}_{qa}^{(2)*} (\Omega_{L,D},0) \mathcal{D}_{qa'}^{(2)} (\Omega_{L,D},t)  \rangle \mathcal{D}_{a0}^{(2)*} (\Omega_{D,i}) \mathcal{D}_{a'0}^{(2)} (\Omega_{D,j}),
	\end{aligned}
\end{equation}
where $D$ is the frame associated to the diffusion tensor, fixed in the molecular frame, $\Omega_{L,D}$ is the time-dependent Euler angle set for the orientation of the diffusion tensor in the laboratory frame and $\Omega_{D,k}$, $k=\{i,j\}$, are the time-independent Euler angle sets for orientation of the interactions $i$ and $j$ in the diffusion frame, fixed in the molecular frame. In the following, we define:
\begin{equation}
C_{aa'} (t) = \langle  \mathcal{D}_{qa}^{(2)*} (\Omega_{L,D},0) \mathcal{D}_{qa'}^{(2)} (\Omega_{L,D},t) \rangle.
\label{eq:CorrFuncTumbInt}
\end{equation}
\paragraph*{Master equation for overall diffusion.} The Master equation for global tumbling, written in a frame where the rotational diffusion tensor is diagonal, is:
\begin{equation}
\frac{\partial}{\partial t} P(\Omega_{L,D}, t) = - \sum_{j=x,y,z} D_{jj} L_j^2 P(\Omega_{L,D},t),
\label{eq:DiffMasterEq}
\end{equation}
where $D_{jj}$ is the $j$th component of the diagonalized diffusion tensor and $L_j$ is the associated angular momentum operators. We express the diffusion operator in terms of raising and lowering operators:
\begin{equation}
\mathcal{D}_{g} = \sum_{j=x,y,z} D_{jj} L_j^2 = D_+ L^2 + \frac{1}{2} D_- (L_+^2 + L_-^2)   + (D_{zz} - D_+)L_z^2,
\label{eq:DtumbExpand}
\end{equation}
where:
\begin{equation}
	\begin{aligned}
	D_\pm =& \frac{1}{2} (D_{xx} \pm D_{yy}),\\
	L_\pm =& L_x \pm i L_y.
	\end{aligned}
\end{equation}
The eigen-equation is written:
\begin{equation}
\mathcal{D}_{g} \Psi_n = E_n \Psi_n,
\label{eq:TumbEigEq}
\end{equation}
where the functions $\Psi_n$ are expanded in terms of the normalized Wigner matrices:
\begin{equation}
\Psi_{L,\mu,\kappa} (\Omega) = \sum_k a_{\kappa,k} \sqrt{\frac{2L+1}{8\pi^2}} \mathcal{D}_{\mu,k}^L (\Omega).
\label{eq:PsiLinComb}
\end{equation}
\paragraph*{Solving the eigen-equation.} The correlation function (Eq.\,\ref{eq:CorrFuncTumbInt}) is expressed as:
\begin{equation}
	\begin{aligned}
	C_{aa'} =& \int d \Omega_0 \int d \Omega  P(\Omega_0) \mathcal{D}_{qa}^{(2)*} (\Omega_0) \mathcal{D}_{qa'}^{(2)} (\Omega) \times \\
		 &\sum_{L,\mu,k, k',\kappa} \frac{2L+1}{8 \pi^2} a_{\kappa, k} a_{\kappa, k'}  \mathcal{D}_{\mu,k}^{L*} (\Omega_0) \mathcal{D}_{\mu,k'}^{L} (\Omega) e^{-E_{L,\mu,\kappa} t}.
	\end{aligned}
	\label{eq:IntegGlobalTumb}
\end{equation}
The orthogonality condition of the Wigner matrices (Eq.\,\ref{eq:PropOrtho}) imposes $L=2$ when calculating the integral. As a consequence, only 5 non-degenerate eigenfunctions are needed to solve Eq.\,\ref{eq:DiffMasterEq}. With this simplification in hand, the eigenvalues and eigenfunctions of the components of the angular momentum operators are:
\begin{equation}
	\begin{aligned}
	L^2 \mathcal{D}_{\mu,k}^{(2)} (\Omega) &= 6 \mathcal{D}_{\mu,k}^{(2)} (\Omega), \\
	L_z^2 \mathcal{D}_{\mu,k}^{(2)} (\Omega) &= k^2 \mathcal{D}_{\mu,k}^{(2)} (\Omega).
	\end{aligned}
	\label{eq:EigMomentumOp}
\end{equation}
\begin{table}[t]
	\caption{Effect of the raising ($L_+$) and lowering ($L_-$) squared operators on the rank-2 Wigner matrices.}%
		{\def\arraystretch{1.8}
		\begin{tabular}{c|cc}
		 	k & $L_+^2 \mathcal{D}_{\mu,k}^{(2)} (\Omega)$ &  $L_-^2 \mathcal{D}_{\mu,k}^{(2)} (\Omega)$ \\
			\hline
			-2 & $\sqrt{24} \mathcal{D}_{\mu,0}^{(2)} (\Omega)$ & 0 \\
			-1 & $6 \mathcal{D}_{\mu,1}^{(2)} (\Omega)$ & 0 \\
			0 & $\sqrt{24} \mathcal{D}_{\mu,2}^{(2)} (\Omega)$ & $\sqrt{24} \mathcal{D}_{\mu,-2}^{(2)} (\Omega)$ \\
			1 & 0 & $6 \mathcal{D}_{\mu,-1}^{(2)} (\Omega)$ \\
			2 & 0 & $\sqrt{24} \mathcal{D}_{\mu,0}^{(2)} (\Omega)$
			\end{tabular}
		}
	\label{table:RaisingLowering}
\end{table}
The effects of the raising and lowering squared operators are summed up in Table\,\ref{table:RaisingLowering}. It is then straightforward to find 3 eigenvectors and associated eigenvalue of $\mathcal{D}_g$ ($\kappa=1,2,3$ in Table\,\ref{table:tumbEigVal}), and the remaining two must have the form: $X(x_1,x_2) = x_1 \sqrt{\frac{5}{8 \pi^2}}\left(\mathcal{D}_{\mu,2}^{(2)} (\Omega) + \mathcal{D}_{\mu,-2}^{(2)} (\Omega) \right) + x_2 \sqrt{\frac{5}{8 \pi^2}}\mathcal{D}_{\mu,0}^{(2)} (\Omega)$, where $x_1, x_2 \in \Re$. The effect of the operator $\mathcal{D}_{g}$ on this eigenfunction is:
\begin{equation}
	\begin{aligned}
\mathcal{D}_{g} X(x_1, x_2) =  &x_1 \sqrt{\frac{5}{8 \pi^2}}\left(\mathcal{D}_{\mu,2}^{(2)} (\Omega) + \mathcal{D}_{\mu,-2}^{(2)} (\Omega) \right) \left[ 6 D_+ + 4(D_{zz} - D_+) + \frac{\sqrt{24}}{2} D_- \frac{x_2}{x_1} \right]  \\
		+& x_2 \sqrt{\frac{5}{8 \pi^2}} \mathcal{D}_{\mu,0}^{(2)} (\Omega) \left[ 6 D_+ + \sqrt{24} D_- \frac{x_1}{x_2} \right],
\label{eq:EigenEquationX}
	\end{aligned}
\end{equation}
which yields to the following condition for $X(x_1, x_2)$ to be an eigenvector:
\begin{equation}
6 D_+ + 4 (D_{zz} - D_+) + \frac{\sqrt{24}}{2} D_- \frac{x_2}{x_1} = 6 D_+ + \sqrt{24} D_- \frac{x_1}{x_2}.
\label{eq:PolyDiffRot}
\end{equation}
In addition, the normalization condition for $X(x_1, x_2)$ gives:
\begin{equation}
2 x_1^2 + x_2^2 = 1.
\end{equation}
We set $x_1'^2 = 2 x_1^2$ such that:
\begin{equation}
x_1'^2 + x_2^2 = 1.
\end{equation}
We can find a number $x$ leading to:
\begin{equation}
		x_1' = \sin(\frac{x}{2}) ,\,\,\,
		x_2 = \cos(\frac{x}{2}).
\end{equation}
After inserting in Eq.\,\ref{eq:PolyDiffRot} and defining $\alpha = \frac{ \sqrt{3} D_-}{D_{zz} - D_+}$, we obtain:
\begin{equation}
x = - \tan^{-1}(\alpha),
\end{equation}
and one set of solution to Eq.\,\ref{eq:PolyDiffRot}:
\begin{equation}
		x_1 = -\frac{1}{\sqrt{2}}\sin \left( \frac{\tan^{-1}(\alpha)}{2} \right) ,\,\,\,
		x_2 = \cos \left( \frac{\tan^{-1}(\alpha)}{2} \right).
\end{equation}
A second set of solution is obtained by choosing:
\begin{equation}
		x_1' = -\cos(\frac{x}{2}) , \,\,\,
		x_2' = \sin(\frac{x}{2}),
\end{equation}
which is a $\frac{\pi}{2}$-shift of the previous solution, and leads to:
\begin{equation}
		x_1' = \frac{1}{\sqrt{2}} \cos \left( \frac{\tan^{-1}(\alpha)}{2} \right) , \,\,\,
		x_2' =  \sin \left( \frac{\tan^{-1}(\alpha)}{2} \right).
\end{equation}
The eigenvalues associated to these two eigenvectors can be calculated easily using Eq.\,\ref{eq:EigenEquationX}. The five eigenvectors and associated eigenvalues of Eq.\,\ref{eq:TumbEigEq} are gathered in Table\,\ref{table:tumbEigVal} where we have defined $\beta = \tan ^{-1} \frac{\sqrt{3} D_-}{D_{zz} - D_+}$. 
\begin{table*}[t]
	\caption{Eigenvalues and eigenvectors for the operator associated to global tumbling. We define $\beta = \tan ^{-1} \frac{\sqrt{3} D_-}{D_{zz} - D_+}$.}
		{\def\arraystretch{1.8}
		\begin{tabular}{c|cc}
		 	$\kappa$ & $E_\kappa$ &  $\Psi_{2,\mu,\kappa} (\Omega)/\sqrt{\frac{5}{8 \pi^2}} = \sum_k a_{\kappa,k} \mathcal{D}_{\mu,k}^{(2)} (\Omega)$ \\
			\hline
			1 & $5D_+ + 3D_- + D_{zz}$ & $\frac{1}{\sqrt{2}}\left(\mathcal{D}_{\mu,1}^{(2)} (\Omega) + \mathcal{D}_{\mu,-1}^{(2)} (\Omega) \right)$ \\
			2 & $5D_+ - 3D_- + D_{zz}$ & $\frac{1}{\sqrt{2}}\left(\mathcal{D}_{\mu,1}^{(2)} (\Omega) - \mathcal{D}_{\mu,-1}^{(2)} (\Omega) \right)$ \\
			3 & $2 D_+ + 4D_{zz}$ & $\frac{1}{\sqrt{2}}\left(\mathcal{D}_{\mu,2}^{(2)} (\Omega) - \mathcal{D}_{\mu,-2}^{(2)} (\Omega) \right)$ \\
			4 & $4D_+ + 2D_{zz} + 2 \left[(D_{zz}-D_+)^2 + 3 D_-^2 \right]^{\frac{1}{2}}$ & $ \frac{1}{\sqrt{2}} \left(\mathcal{D}_{\mu,2}^{(2)} (\Omega) + \mathcal{D}_{\mu,-2}^{(2)} (\Omega) \right) \cos \frac{\beta}{2} + \mathcal{D}_{\mu,0}^{(2)} (\Omega) \sin \frac{\beta}{2}$ \\
			5 & $4D_+ + 2D_{zz} - 2 \left[(D_{zz}-D_+)^2 + 3 D_-^2\right]^{\frac{1}{2}}$ & $- \frac{1}{\sqrt{2}} \left(\mathcal{D}_{\mu,2}^{(2)} (\Omega) + \mathcal{D}_{\mu,-2}^{(2)} (\Omega) \right) \sin \frac{\beta}{2} + \mathcal{D}_{\mu,0}^{(2)} (\Omega) \cos \frac{\beta}{2}$ 
		\end{tabular}
		}
	\label{table:tumbEigVal}
\end{table*}
When the diffusion tensor shows some symmetry properties, the angular momentum operator can have degenerate eigenfunctions. For example, if the molecule behaves as a cylinder (axial symmetry), then $D_{xx}=D_{yy}$ (a frame transformation can always lead to this situation), eliminating the $L_+$ and $L_-$ parts of Eq.\,\ref{eq:DtumbExpand}. The immediate consequence is that $D_-=0$ and two of the three unique eigenvalues are doubly degenerate ($\kappa=1$ and 2, and $\kappa=3$ and 4 for $D_z > D_x$ or $\kappa=3$ and 5 for $D_z < D_x$). When the diffusion tensor is isotropic, $D_{xx}=D_{yy}=D_{zz}=D$, all eigenvalues are degenerate and equal $6D$, usually refered to as the inverse of the global tumbling correlation time: $\tau_c = 1/(6D)$. 
\paragraph*{Static angle probability distribution.} It is clear from the form of the eigenvalues of the components of the angular momentum that one only needs to calculate the eigenfunction associated with the 0 eigenvalue in order to have a non-vanishing exponential term in Eq.\,\ref{eq:PropCondProb} when calculating the limit with $t \rightarrow \infty$. It comes, from the eigenvalues of $L^2$ (Eq.\,\ref{eq:EigMomentumOp}), that such an eigenfunction has $L=0$ and, from the eigenvalues of $L_z$, $k=0$ such that it corresponds to $D_{0, 0}^{(0)}(\Omega)$. Then:
\begin{equation}
P(\Omega_0) = \frac{1}{8 \pi^2} D_{0,0}^{(0)}(\Omega_0) ^2 = \frac{1}{8 \pi^2}.
\end{equation}
\paragraph*{Integration of the correlation function.}
Using the integration in Eq.\,\ref{eq:IntegGlobalTumb}:
\begin{equation}
	\begin{aligned}
	C_{aa'} (t) =& \int d \Omega_0 \int d \Omega \sum_{\mu,k,k'=-2}^2 \sum_{\kappa=1}^5  P(\Omega_0)  \frac{5}{8 \pi^2} \times  \\
		&e^{-E_\kappa t} a_{\kappa,k} a_{\kappa,k'} \mathcal{D}_{qa}^{(2)*}(\Omega_0) \mathcal{D}_{\mu k}^{(2)*}(\Omega_0)  \mathcal{D}_{qa'}^{(2)}(\Omega) \mathcal{D}_{\mu k'}^{(2)}(\Omega).
	\end{aligned}
\end{equation}
We can use Eq.\,\ref{eq:PropConjWigner} to transform this equation into:
\begin{equation}
	\begin{aligned}
	C_{aa'} (t) =&  \frac{5}{64 \pi^4}  \sum_{\mu, k, k'=-2}^2 \sum_{\kappa=1}^5 a_{\kappa,k} a_{\kappa,k'} e^{-E_\kappa t} \times \\
		&  \times  (-1)^{-k} \int d \Omega_0 \mathcal{D}_{qa}^{(2)*}(\Omega_0) \mathcal{D}_{-\mu -k}^{(2)}(\Omega_0) \times \\
		 &\times (-1)^{-k'} \int d \Omega  \mathcal{D}_{qa'}^{(2)}(\Omega) \mathcal{D}_{-\mu -k'}^{(2)*}(\Omega),
	\end{aligned}
\end{equation}
According to Table\,\ref{table:tumbEigVal}, $k+k'$ is always an even number such that $(-1)^{-k-k'}=1$ for all $k$ and $k'$. It immediately follows from the orthonormality condition (Eq.\,\ref{eq:PropOrtho}) that:
\begin{equation}
	C_{aa'} (t) = \frac{5}{64 \pi^4} \sum_{\mu, k, k'=-2}^2 \sum_{\kappa=1}^5 \delta_{q-\mu} \delta_{a-k} \delta_{a'-k'}   a_{\kappa,k} a_{\kappa,k'} e^{-E_\kappa t}  \frac{8 \pi^2}{5} \times \frac{8 \pi^2}{5},
\end{equation}
where $\delta$ is the Kronecker-delta function. It simplifies into:
\begin{equation}
C_{aa'} (t) = \frac{1}{5}  \sum_{\kappa=1}^5 a_{\kappa,-a} a_{\kappa,-a'} e^{-E_\kappa t}\,.
\end{equation}
From symmetry consideration, it follows that $a_{\kappa,-a} a_{\kappa,-a'}=a_{\kappa,a} a_{\kappa,a'}$ so that:
\begin{equation*}
C_{aa'} (t) = \frac{1}{5}  \sum_{\kappa=1}^5 a_{\kappa,a} a_{\kappa,a'} e^{-E_\kappa t},
\end{equation*}
and the correlation function can now be written as:
\begin{equation}
C_{i,j} (t) = \frac{1}{5} \sum_{\kappa=1}^5  \sum_{a, a'=-2}^2 a_{\kappa,a} a_{\kappa,a'} e^{-E_\kappa t}   \mathcal{D}_{a,0}^{(2)*} (\Omega_{D,i}) \mathcal{D}_{a',0}^{(2)} (\Omega_{D,j}).
\label{eq:FullCorrTumblingOnly}
\end{equation}
\paragraph*{Special case 1: axial symmetry.}
In this case, the Wigner matrices $\mathcal{D}_{\mu,k}^{(2)}(\Omega), k \in [-2, 2]$ are eigenfunctions of the diffusion operator with associated eigenvalue $6 D_\perp + k^2 \left( D_\parallel - D_\perp \right)$ where we define $D_\perp = D_{xx} = D_{yy}$ and $D_\parallel = D_{zz}$. It follows:
\begin{equation}
C_{aa'} (t) = \delta_{aa'} \frac{1}{5} e^{-(6 D_\perp + a^2 (D_\parallel - D_\perp)) t},
\end{equation}
and the total correlation function is:
\begin{equation}
C_{i,j} (t) = \frac{1}{5} \sum_{a=-2}^2 e^{-(6 D_\perp + a^2 (D_\parallel - D_\perp)) t} \mathcal{D}_{a,0}^{(2)*} (\Omega_{D,i}) \mathcal{D}_{a,0}^{(2)} (\Omega_{D,j}).
\label{eq:FullCorrSymTopTumblingOnly}
\end{equation}
\paragraph*{Special case 2: isotropic tumbling.}
In this case as well, the Wigner matrices $\mathcal{D}_{\mu,k}^{(2)}(\Omega), k \in [-2, 2]$ are eigenfunctions of the diffusion operator with degenerate eigenvalue of $6D$. The well-known result immediately follows:
\begin{equation}
C_{aa'} (t) = \delta_{aa'} \frac{1}{5}  e^{-t/\tau_c},
\end{equation}
where the global tumbling correlation time is defined as:
\begin{equation}
\tau_c = \frac{1}{6 D }.
\end{equation}
The total correlation function is:
\begin{equation}
C_{i,j} (t) = \frac{1}{5} \sum_{a=-2}^2 e^{-t/\tau_c} \mathcal{D}_{a,0}^{(2)*} (\Omega_{D,i}) \mathcal{D}_{a,0}^{(2)} (\Omega_{D,j}),
\end{equation}
and after using Eq.\,\ref{eq:SumProdWignerP2}, we have:
\begin{equation}
C_{i,j} (t) = \frac{1}{5} e^{-t/\tau_c} \mathcal{P}_2 (\cos \theta_{i,j}),
\label{eq:FullCorrSphereTopTumblingOnly}
\end{equation}
where $\theta_{i,j}$ is the angle between interactions $i$ and $j$. In the case of small anisotropy of the overall diffusion tensor, this simpler model of correlation function can be used with residue-specific correlation time for global tumbling defined from the projection of the diffusion tensor onto the interaction frame \cite{Bruchweiler_Science_1995}.

\subsection{Rotamer jumps}
In this section, we add one internal motion on top of the global tumbling. We assume that the moiety in which the principal axis frame of the interactions is anchored (e.g. a peptide plane, or methyl group) can adopt a finite number of fixed conformations and that this moiety can exchange between all conformations. Moreover, it is assumed that the transition events are infinitely fast, that is, the system can always be found in one particular conformation. This model has been used to describe methyl rotation as a three-site jump\cite{batchelder_jacs_1983,vugmeyster_methods_2018} and is well adapted to study side-chain motions which can undergo fast rotamer transitions, as revealed by MD simulations \cite{Best_JACS_2004,Cousin_JACS_2018,Kummerer_JCTP_2020,Hoffmann_JPCB_2022}. \\
The treatment of rotameric jumps in this section is based on the work of Wittebort and Szabo \cite{Witterbort_JCP_1978}. We will use the result from the previous section and write the correlation function as:
\begin{equation}
C_{i,j} (t) = \frac{1}{5} \sum_{\kappa=1}^5  \sum_{a, a'=-2}^2  a_{\kappa,a} a_{\kappa,a'} e^{-E_\kappa t}  \langle \mathcal{D}_{a,0}^{(2)*} (\Omega_{D,i}, 0) \mathcal{D}_{a',0}^{(2)} (\Omega_{D,j}, t) \rangle ,
\end{equation}
where the ensemble average $\langle ... \rangle$ accounts for the presence of internal motions. This equation is valid only under the assumption that global tumbling and internal motions are uncorrelated. This allows to introduce the jump frame for a facilitated description of the jump. For example, in the case of valines $\chi_1$ rotameric jumps (corresponding to different orientations of the C$_{\beta}$-C$_{\gamma1}$ axis), the jump frame main axis ($z_J$) points along the C$_\alpha$-C$_\beta$  bond, with origin at the C$_\beta$. The orientation of $x_J$ axis is arbitrary. The Wigner matrices can be split into successive frame transformations:
\begin{itemize}
\item from the diffusion frame to the jump frame, with Euler angle $\Omega_{D,J}$. This frame is fixed in the diffusion frame, and the transformation is time-independent.
\item from the jump frame to  the system frame (SF), with Euler angle $\Omega_{J,SF}= \{\alpha_{J,SF}, \beta_{J,SF}, 0 \}$. The main axis of the system frame points along the chemical bond defining the conformation. This transformation is time-dependent.
\item from the system frame to the principal axis frame (PAF) of the interaction. For instance, the PAF of a dipole-dipole interaction has its main axis pointing along the internuclear vector. This is a time-independent transformation, with Euler angle $\Omega_{SF, i}$. 
\end{itemize}
These additional frame transformations lead to:
\begin{equation}
	\begin{aligned}
C_{i,j} (t) = &\frac{1}{5}  \sum_{\kappa=1}^5  \sum_{a,a',b,b',c,c'=-2}^2  a_{\kappa,a} a_{\kappa,a'} e^{-E_\kappa t} \langle \mathcal{D}_{b,c}^{(2)*} (\Omega_{J,SF}, 0)  \mathcal{D}_{b',c'}^{(2)} (\Omega_{J,SF}, t) \rangle \times \\
	&  \mathcal{D}_{a,b}^{(2)*} (\Omega_{D,J}) \mathcal{D}_{a',b'}^{(2)} (\Omega_{D,J}) \mathcal{D}_{c, 0}^{(2)*} (\Omega_{SF, i}) \mathcal{D}_{c', 0}^{(2)} (\Omega_{SF, j}).
\label{eq:CorrJumpTotPre}
	\end{aligned}
\end{equation}
The conditional probability $P(\Omega, t | \Omega_0, 0)$ (Eq.\,\ref{eq:PropCondProb}) is found by solving the Master equation:
\begin{equation}
\frac{\partial}{\partial t}p_\alpha (t)  = \sum_{\beta=1}^N \mathcal{R}_{\alpha \beta} p_\beta (t),
\label{eq:MasterEqJump}
\end{equation}
where $p_\alpha (t)$ is the population of state $\alpha$ at time $t$ and $\mathcal{R}_{\alpha \beta}$ is an element of the exchange matrix $\mathcal{R}$ and corresponds to the exchange rate from state $\beta$ to $\alpha$. The microscopic reversibility implies that:
\begin{equation}
\mathcal{R}_{\alpha  \beta} p_\beta^{eq} = \mathcal{R}_{\beta \alpha} p_\alpha^{eq},
\label{eq:MicroRev}
\end{equation}
and diagonal elements of the exchange matrix are given by:
\begin{equation}
\mathcal{R}_{\alpha \alpha} = - \sum_{\beta \neq \alpha} R_{\beta \alpha}.
\end{equation}
The boundary conditions associated to Eq.\,\ref{eq:MasterEqJump} are:
\begin{equation}
	\begin{aligned}
	&(1)\,\,\, p(\beta, t=0 | \alpha, 0) = \delta_{\alpha \beta}, \\
	&(2)\,\,\, \lim_{t \rightarrow \infty} p(\beta, t | \alpha, 0)  = p_\beta^{eq},
	\end{aligned}
\label{eq:ConditionJumpsProb}
\end{equation}
Eq.\,\ref{eq:MasterEqJump} transforms into an eigen-equation as follows:
\begin{equation}
\mathcal{R} X = \lambda X.
\end{equation}
The exchange matrix is in general not symmetric (it is only when equilibrium populations are equal) but can be transformed into a symmetric matrix $\tilde{\mathcal{R}}$ with coefficients:
\begin{equation}
	\begin{aligned}
	\widetilde{\mathcal{R}}_{\alpha \beta} &= \sqrt{\mathcal{R}_{\alpha \beta} \mathcal{R}_{\beta \alpha}},\,\,\, \alpha \neq \beta, \\
	\widetilde{\mathcal{R}}_{\alpha \alpha} &= \mathcal{R}_{\alpha \alpha}.
	\end{aligned}
\label{eq:SymExMatrixDefinition}
\end{equation}
The eigen-equation now reads:
\begin{equation}
\tilde{\mathcal{R}} \tilde{X} = \lambda \tilde{X},
\end{equation}
where the $\alpha^\mathrm{th}$ coordinate of the eigenvector associated to the eigenvalue $\lambda_n$ is:
\begin{equation}
\tilde{X}_\alpha^{(n)} = \frac{1}{\sqrt{p_\alpha^{eq}}} X_\alpha^{(n)}.
\end{equation}
Since $\tilde{\mathcal{R}}$ is symmetric, its eigenvalues $\lambda_n$ are all real and the associated eigenvectors $\tilde{X}^{(n)}$ are orthogonal. The conditional probability is written as in Eq.\,\ref{eq:PropCondProb} such that, for the second condition in Eq.\,\ref{eq:ConditionJumpsProb}  to be  met, we must have $\lambda_0 = 0$ (we order the eigenvalues such that $\left| \lambda_n \right | \leq \left | \lambda_{n+1} \right| $) and the associated eigenvector is $\{\tilde{X}_n^{(0)}\}  = \{\sqrt{p_n^{eq}}\}$. \\
The conditional probability can be written:
\begin{equation}
p(\beta, t | \alpha, 0)  = \sqrt{\frac{p_\beta^{eq}}{p_\alpha^{eq}}} \sum_{n = 0}^{N-1} \tilde{X}_\alpha ^{(n)} \tilde{X}_\beta ^{(n)} e^{\lambda_n  t},
\end{equation}
where the factor $\sqrt{\frac{p_\beta^{eq}}{p_\alpha^{eq}}}$ is introduced to fulfill the second condition in Eq.\,\ref{eq:ConditionJumpsProb} (the eigenvectors $\tilde{X}^{(n)}$ are orthogonal such that condition (1) is already met). It leads to the following expression for the correlation function (note that integrals in Eq.\,\ref{eq:PropIntegrationC} are replaced by  discrete sums):
\begin{equation}
\langle \mathcal{D}_{b,c}^{(2)*} (\Omega_{J,SF}, 0)  \mathcal{D}_{b',c'}^{(2)} (\Omega_{J,SF}, t) \rangle = \sum_{\alpha,\beta=1}^N \sum_{n=0}^{N-1} e^{\lambda_n  t}  \sqrt{p_\alpha^{eq}p_\beta^{eq}}\tilde{X}_\alpha ^{(n)} \tilde{X}_\beta ^{(n)} \mathcal{D}_{b,c}^{(2)*} (\Omega_{J,SF_\alpha}) \mathcal{D}_{b',c'}^{(2)} (\Omega_{J,SF_\beta}).
\label{eq:RotJumCorrFunc}
\end{equation}
The total correlation function accounting for overall rotational diffusion and rotamer jumps is then written:
\begin{equation}
C_{i,j} (t) = \frac{1}{5} \sum_{\kappa=1}^5  \sum_{a,a',=-2}^2 \sum_{\alpha,\beta=1}^ N \sum_{n=0}^{N-1} a_{\kappa,a} a_{\kappa,a'}    e^{-E_\kappa t}    \sqrt{p_\alpha^{eq}p_\beta^{eq}} \tilde{X}_\alpha ^{(n)} \tilde{X}_\beta ^{(n)} e^{\lambda_n  t}  \mathcal{D}_{a,0}^{(2)*} (\Omega_{D,i_\alpha}) \mathcal{D}_{a',0}^{(2)} (\Omega_{D,j_\beta}),
\label{eq:JumpsCompact}
\end{equation}
where $\Omega_{D,i_\alpha}$ is the Euler angle set for transformation from the diffusion frame to the interaction-$i$ frame in rotamer $\alpha$. Thus, in addition to report on the kinetics of the exchange (populations and jump rates), the correlation function for rotamer jump also depends on the resulting distinct orientations of the interaction frames in the diffusion frame. This can be particularly critical when the overal diffusion tensor is anisotropic, as discussed below.
\paragraph*{Order parameter for rotamer jumps.}
The internal correlation function for rotamer jumps in Eq.\,\ref{eq:RotJumCorrFunc} does not cancel out only for $n = 0$ when $t \rightarrow \infty$, so that the squared order parameter for rotamer jumps is, after using Eq.\,\ref{eq:SumProdWignerP2} \cite{Chou_JACS_2003}:
\begin{equation}
\mathcal{S}_{J}^2(i,j) = \sum_{\alpha=1}^N \sum_{\beta=1}^N   p_\alpha^{eq}p_\beta^{eq}\mathcal{P}_2 (\cos \theta_{i_\alpha, j_\beta}),
\label{eq:OrderParamJumps}
\end{equation}
where $\theta_{i_\alpha, j_\beta}$ is the angle between interaction $i$ in rotamer $\alpha$ and interaction $j$ in rotamer $\beta$.

\subsection{Rotation on a cone}
The diffusion on a cone can be used, for example, to model the rotation of a methyl group around its symmetry axis under the assumption that thermal energy is much larger than the energy barrier for the rotation, so that the probability distribution for the position of hydrogen atoms is uniform. Methyl groups are now widely used in biomolecular NMR for their favorable relaxation properties that precisely originate from the fast rotation around their symmetry axis \cite{Tugarinov_JACS_2003, Tugarinov_NatProtoc_2006,Gans_ANIE_2010} .In particular, transverse-relaxation optimized spectroscopy (TROSY) of methyl groups has made it possible to investigate large proteins and protein complexes. This approach is based on the cancelation between proton-proton dipolar auto- and cross-relaxation mechanisms \cite{Tugarinov_JACS_2003,Ollerenshaw_MRC_2003} for which fast methyl rotation is essential \cite{BolikCoulon_JCP_2019}. In this context, force fields used in MD simulations of proteins have been modified recently using NMR experimental data to correctly account for the energy barrier for the rotation \cite{Hoffmann_JPhysChemB_2018,Hoffmann_JCP_2020}. The solution of the Master equation (Eq.\,\ref{eq:CorrFuncApproachOp}) has first been published by D.\,Wallach \cite{Wallach_JCP_1967}. \\
In order to describe the diffusion on a cone motion, an additional frame needs to be introduced on top of the diffusion and interaction frames. Similarly to the rotamer jumps, we call this frame the system frame (SF). Its main axis is pointing along the axis of rotation, with the x- and y-axis rotating such that the transformation from the SF to the interaction frame(s) is time-independent. The Euler angle for transformation from the diffusion frame to the system frame is $\Omega_{D,SF} = \{\alpha_{D,SF}, \beta_{D,SF}, \gamma_{D,SF}(t) \}$ where only the third angle is time-dependent. The correlation function is then:
\begin{equation}
	\begin{aligned}
C_{i,j} (t) =& \frac{1}{5} \sum_{\kappa=1}^5 \sum_{a,a',b,b'=-2}^2 a_{\kappa,a} a_{\kappa,a'} e^{-E_\kappa t} e^{i \alpha_{D,SF} (a - a')} d_{a,b} (\beta_{D,SF}) d_{a',b'} (\beta_{D,SF}) \times \\
		& \langle e^{i (b\gamma_{D,SF}(0) - b'\gamma_{D,SF} (t))} \rangle  \mathcal{D}_{b,0}^{(2)*} (\Omega_{SF,i}) \mathcal{D}_{b',0}^{(2)} (\Omega_{SF,j}).
	\end{aligned}
\label{eq:CorrFuncRotPre}
\end{equation}
The Master equation is written using the angular momentum operator $L_{rot}^2$:
\begin{equation}
\frac{\partial}{\partial t} p(\gamma, t) = - D_{rot} L_{rot}^2 p(\gamma, t) = D_{rot} \frac{\partial^2}{\partial \gamma^2} p(\gamma, t),
\label{eq:MasterEqRot}
\end{equation}
where $D_{rot}$ is the rotational diffusion coefficient. The ensemble average is calculated as:
\begin{equation}
\langle e^{i (b\gamma_{D,SF}(0) - b'\gamma_{D,SF} (t))} \rangle = \int_0^{2\pi} \int_0^{2\pi}  e^{i b \gamma_0} e^{-i b' \gamma}   p(\gamma_0) p(\gamma, t | \gamma_0, 0) \mathrm{d} \gamma_0 \mathrm{d} \gamma,
\label{eq:GeneralCorrRotCone}
\end{equation}
where $p(\gamma_0) = \frac{1}{2\pi}$ (all orientations are equi-probable), and the conditional probability is expressed as:
\begin{equation}
p(\gamma, t | \gamma_0, 0) = \sum_n \varphi_n^* (\gamma_0) \varphi_n (\gamma) e^{-\lambda_n t},
\end{equation}
where $\varphi_n$ is the eigenfunction associated to the eigenvalue $\lambda_n$ of the rotation operator $D_{rot} L_{rot}^2$. Eigenfunctions and eigenvalues are given by:
\begin{equation}
	\begin{aligned}
&\varphi_n (\gamma) = \frac{1}{\sqrt{2\pi}} e^{i n \gamma}, \\
&\lambda_n = D_{rot} n^2.
	\end{aligned}
\end{equation}
Inserting in Eq.\,\ref{eq:GeneralCorrRotCone} leads to:
\begin{equation}
 \langle e^{i (b\gamma_{D,SF}(0) - b'\gamma_{D,SF} (t))} \rangle = \frac{1}{4\pi^2} \sum_n e^{-D_{rot} n^2 t}\int_0^{2\pi} \int_0^{2\pi} e^{i \gamma_0 (b - n)} e^{-i \gamma(b'- n)} \mathrm{d}\gamma_0 \mathrm{d}\gamma, 
\end{equation}
which does not cancel out only for $n=b=b'$, and leads to the correlation function:
\begin{equation}
\langle e^{i (b\gamma_{D,SF}(0) - b'\gamma_{D,SF} (t))} \rangle = \delta_{bb'} e^{-D_{rot} b^2 t}.
\label{eq:RotConeCorrFunc}
\end{equation}
The total correlation function is then:
\begin{equation}
	\begin{aligned}
 C_{i,j}(t) =& \frac{1}{5} \sum_{\kappa=1}^5 \sum_{a,a',b=-2}^2 a_{\kappa,a} a_{\kappa,a'} e^{-E_\kappa t} e^{-D_{rot} b^2 t} \times \\
		& e^{i \alpha_{D,SF} (a - a')} d_{a,b} (\beta_{D,SF}) d_{a',b} (\beta_{D,SF}) \mathcal{D}_{b,0}^{(2)*} (\Omega_{SF,i}) \mathcal{D}_{b,0}^{(2)} (\Omega_{SF,j}).
	\end{aligned}
\end{equation}
\paragraph*{Order parameter for rotation on a cone.}
The only non-vanishing term in Eq.\,\ref{eq:RotConeCorrFunc}, when $t \rightarrow \infty$, is obtained for $b=0$, so that the squared order-parameter for rotation on a cone is:
\begin{equation}
\mathcal{S}_{R}^2 (i,j) = \mathcal{P}_2 (\cos \beta_{SF,i}) \mathcal{P}_2 (\cos \beta_{SF,j}).
\end{equation}
In the case of methyl-group rotation, with ideal tetrahedral geometry, it leads to the well-known result $\mathcal{S}_{met}^2 (\mathrm{DD}_\mathrm{CH},\mathrm{DD}_\mathrm{CH})=1/9$, where $\mathrm{DD}_\mathrm{CH}$ stands for the C-H dipole-dipole (DD) interaction.
\paragraph*{CSA/DD cross-correlation.}
\begin{figure*}[!ht]
		\includegraphics[width=0.9\textwidth]{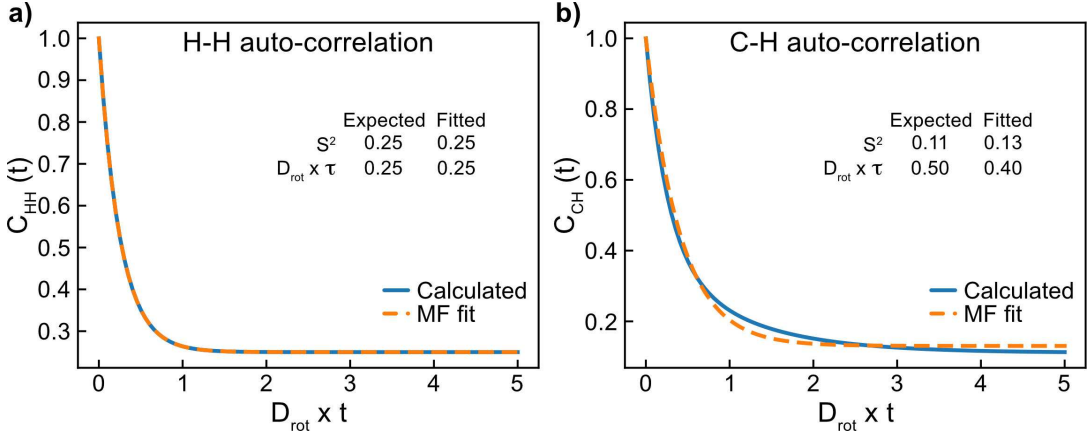}
	\caption{Correlation function for methyl group rotation. Auto-correlation function for H-H  (\textbf{a}) and C-H (\textbf{b}) motions. Calculated correlation functions are shown in blue, and MF fits in dash orange. The  correlation functions are shown as a function of $D_{rot} \times t$ such that they are independent of the diffusion constant $D_{rot}$. The expected and fitted MF parameters are shown on each pannel.}
	\label{fig:MFmetRot}
\end{figure*}
In methyl groups, carbon-13 Chemical Shift Anisotropy (CSA) and DD cross-correlated cross-relaxation rates can be measured \cite{Pelupessy_JCP_2007}. In this case, the carbon-13 CSA doesn't undergo a rotation motion and the correlation function in Eq.\,\ref{eq:CorrFuncRotPre} is modified to:
\begin{equation}
	\begin{aligned}
C_\mathrm{CSA,DD} (t) = \frac{1}{5}& \sum_{\kappa=1}^5 \sum_{a,a',b'=-2}^2 a_{\kappa,a} a_{\kappa,a'} e^{-E_\kappa t}  \mathcal{D}_{a,0}^{(2)*} (\Omega_{D,CSA}) e^{i a' \alpha_{D,SF}} d_{a',b'} (\beta_{D,SF}) \times \\
		& \langle e^{i b'\gamma_{D,SF}(t)} \rangle \mathcal{D}_{b',0}^{(2)} (\Omega_{SF,DD}).
	\end{aligned}
\end{equation}
Due to the methyl rotation, $\langle e^{i b'\gamma_{D,SF}(t)} \rangle$ equals 0 on average unless $b'=0$ such that:
\begin{equation}
C_\mathrm{CSA,DD} (t) = \frac{1}{5} \sum_{\kappa=1}^5 \sum_{a,a'=-2}^2 a_{\kappa,a} a_{\kappa,a'} e^{-E_\kappa t} e^{i a' \alpha_{D,SF}}  \mathcal{P}_2(\cos \beta_{SF,DD})\mathcal{D}_{a,0}^{(2)*} (\Omega_{D,CSA}) d_{a',0} (\beta_{D,SF}),
\label{eq:CSADDcrosscorrelation}
\end{equation}
where $\mathcal{P}_2(\cos \beta_{SF,DD})=-1/3$ for methyl groups with perfect tetrahedral geometry.
\paragraph*{C-H and H-H auto-correlation.} The DD contribution to carbon-13 relaxation rates only reports on the C-H correlation, while for protons (or deuterons in the case of specific labeling) it reports on both C-H and H-H correlations. For the sake of simplicity, we assume here an isotropic overall diffusion tensor and a perfect tetrahedral geometry for the methyl group to expand the auto-correlation functions as:
\begin{equation}
	\begin{dcases}
C_{CH}(t) = \frac{1}{5} e^{-t/\tau_c} \left(\frac{1}{9} + \frac{8}{27} e^{- D_{rot} t} + \frac{16}{27} e^{- 4D_{rot} t}  \right), \\
C_{HH} (t) = \frac{1}{5} e^{-t/\tau_c} \left(\frac{1}{4} + \frac{3}{4}  e^{-4 D_{rot} t} \right).
	\end{dcases}
\end{equation}
It is interesting to note that the H-H auto-correlation function is written in the form of the MF correlation function with generalized squared order parameter $\mathcal{S}_{HH}^2 = 1/4$ and correlation time $\tau_{HH} = 1/(4D_{rot})$. The C-H auto-correlation function can be approximated to a single-decay exponential with generalized squared order parameter $\mathcal{S}_{CH}^2 = 1/9$ and correlation time defined to keep the area of the correlation function constant:
\begin{equation}
\tau_{CH} = \frac{1}{1 - \mathcal{S}_{CH}^2} \sum_{b \neq 0} \frac{1}{b^2 D_{rot}} \left[ d_{b,0} (\beta_{met}) \right]^2 = 2 \tau_{HH},
\label{eq:TauCHEffRotMF}
\end{equation}
where $\beta_{met} = 109.47^\circ$. Thus, C-H and H-H auto-correlation lead to a methyl rotation correlation different, in theory, by a factor of 2 when modeled using MF correlation functions. This is shown in Fig.\,\ref{fig:MFmetRot} where the calculated internal correlation function for methyl rotation is perfectly fitted with the MF correlation function for H-H auto-correlation, but show some deviations in the case of C-H auto-correlation with pseudo-correlation time $D_{rot}\times\tau$ significantly different between the two types of correlation despite the fact they report on the same motion. This has no visible effects on the carbon longitudinal and transverse relaxation rates (Fig.\,S1.a,b) and slight deviations in carbon-proton DD cross-relaxation rates at high magnetic-fields (Fig.\,S1.c). \\
When the energy barrier for the rotation is significantly larger than thermal energy \cite{Hoffmann_JPhysChemB_2018,Hoffmann_JCP_2020}, a jump motion between three discrete positions might be more physically relevant \cite{Witterbort_JCP_1978,batchelder_jacs_1983,vugmeyster_methods_2018}. The rotation on a cone model if however mathematically simpler, and for effective correlation time for methyl rotation smaller than 50\,ps, the two models are virtually indistinguishable in the simple case evaluated in Fig.\,S2. When a three-site jump model is considered, one should keep in mind that diffusion within each methyl rotamer energy well is expected to cover a rather wide angle. This can be accounted for by the restricted rotation in a cone model.

\paragraph*{Restricted rotation on a cone}
In the case where the rotation is restricted to a portion of the cone surface, for example and without loss of generality for angles $\varphi_{D,SF}$ between $\pm \gamma_c$, the Master equation Eq.\,\ref{eq:MasterEqRot} has reflecting boundary condition:
\begin{equation}
\frac{\partial p(\gamma, t)}{\partial \gamma}  |_{\gamma=\pm \gamma_c} = 0.
\end{equation}
Solutions for the Master equation have been reported by Wittebort and Szabo \cite{Witterbort_JCP_1978}:
\begin{equation}
	\begin{aligned}
&\varphi_n (\gamma) = \cos \frac{n\pi (\gamma - \gamma_c)}{2 \gamma_c}, \\
&\lambda_n = D_{rot}  \frac{n^2 \pi^2}{4 \gamma_c^2},
	\end{aligned}
\end{equation} 
leading to:
\begin{equation}
	\begin{aligned}
 \langle e^{i (b\gamma_{D,SF}(0) - b'\gamma_{D,SF} (t))} \rangle = \frac{1}{4\gamma_c^2} \sum_n e^{-D_{rot} \frac{n^2 \pi^2}{4 \gamma_c^2} t}  \int_{-\gamma_c}^{+\gamma_c} e^{i \gamma_0 b } \cos \frac{n\pi (\gamma_0 - \gamma_c)}{2 \gamma_c} \mathrm{d}\gamma_0  & \times \\
	 \int_{-\gamma_c}^{+\gamma_c}  e^{-i \gamma b'} \cos \frac{n\pi (\gamma - \gamma_c)}{2 \gamma_c} \mathrm{d}\gamma& , 
	\end{aligned}
\end{equation}
which integrates into:
\begin{equation}
	\begin{aligned}
\langle e^{i (b\gamma_{D,SF}(0) - b'\gamma_{D,SF} (t))} \rangle = \frac{b b' \gamma_c^2}{2(b^2 \gamma_c^2 - \frac{n^2 \pi^2}{4})(b'^2 \gamma_c^2 - \frac{n^2 \pi^2}{4})} \times & \\
\left( \sin b\gamma_c \sin b'\gamma_c (1 + (-1)^n) + \cos b\gamma_c \cos b'\gamma_c (1 - (-1)^n)  \right),
	\end{aligned}
\end{equation}
and leads to the correlation function:
\begin{equation}
	\begin{aligned}
 C_{i,j}(t) = \frac{1}{5} \sum_{\kappa=1}^5 \sum_{a,a',b,b'=-2}^2 \sum_{n=0}^{+\infty} a_{\kappa,a} a_{\kappa,a'} e^{-E_\kappa t} \mathcal{D}_{a,b}^{(2)*}(\Omega_{D,SF})\mathcal{D}_{a',b'}^{(2)}(\Omega_{D,SF}) \mathcal{D}_{b,0}^{(2)*} (\Omega_{SF,i}) \mathcal{D}_{b',0}^{(2)} (\Omega_{SF,j})  \times &\\
		  \frac{bb'\gamma_c^2 e^{-D_{rot} \frac{n^2 \pi^2}{4\gamma_c^2} t} }{2(b^2\gamma_c^2 - \frac{n^2 \pi^2}{4}) (b'^2 \gamma_c^2 - \frac{n^2 \pi^2}{4})}( \sin b\gamma_c \sin b'\gamma_c (1 + (-1)^n) +  \cos b\gamma_c \cos b'\gamma_c (1 - (-1)^n)  ) .
	\end{aligned}
\end{equation}
The sum over $n$ has been reported to converge rapidly \cite{Witterbort_JCP_1978} such that only the first 5 terms can be considered when calculating the correlation function. Restricted rotation on a cone can also be treated using the Gaussian Axial Fluctuation (GAF) model which assumes a Gaussian shape for the probability distribution \cite{Bruschweiler_JACS_1994,Bremi_JACS_1997_2,Bremi_JACS_1997}. Such correlation function to analyse methyl rotation could be combined with a jump-type of model to represent the dynamic as a three-site hope with restricted diffusion within each site. 

\subsection{Wobbling in a cone}
The last type of motion that we will consider here is the wobbling in a cone. In this model, the bond vector undergoes restricted diffusion within a solid angle that is constrained by a cone. It can be useful to model the motion of any bonds such as a $^{15}$N-$^1$H pair, or C-C bonds along protein side-chains. The correlation function for this type of motion was initially introduced in fluorescence \cite{Kinosita_BiophysJ_1977} and dielectric spectroscopy \cite{Warchol_AdvMolRelax_1978}. As will be seen below, the Master equation can be solved analytically, but the complexity of the solution has led to the introduction of approximated forms for the correlation function by the group of A.\,Ikegami \cite{Kinosita_BiophysJ_1977}, and later obtained similarly by G.Lipari and A.\,Szabo \cite{Lipari_BiophysJ_1980,Lipari_JCP_1981}. which guided the latter two towards the MF correlation function \cite{Lipari_JACS_1982_1}. However, the initial treatment of this model and the following studies were performed under the assumption that the interaction of interest was undergoing the wobbling motion, which makes it \textit{a priori} not applicable to the study of motions of a CC bond with $^1$H-$^{13}$C relaxation. A.\,Kumar solved the Master equation when the frame undergoing the diffusion in a cone is not the interaction frame \cite{Kumar_JCP_1986, Kumar_JCP_1989}. We will follow his approach, as well as the one of C.\,Wang and R.\,Pecora \cite{Wang_JCP_1980} who extended the work of M.\,Warchol and W.\,Vaughan \cite{Warchol_AdvMolRelax_1978}. In the construction of this model, the diffusing bond vector also undergoes rotation around itself, which is not likely to correctly reproduce motions of side-chains C-C bond vectors. We introduce here a way to overcome this difficulty. \\
For the description of this motion, we introduce the  Wobbling Frame (WF) which main axis points along the symmetry axis of the cone, and x-axis can point in any direction. Similarly to what was presented in the previous sections, we also include a System Frame (SF) which undergoes the diffusion motion. Assuming the wobbling and global tumbling are uncorrelated, the correlation function is written:
\begin{equation}
	\begin{aligned}
C_{i,j}(t) = &\frac{1}{5} \sum_{\kappa=1}^5 \sum_{a,a',b,b',c,c'=-2}^2  a_{\kappa,a} a_{\kappa,a'} e^{-E_\kappa t}  \langle \mathcal{D}_{b,c}^{(2)*} (\Omega_{WF,SF},0) \mathcal{D}_{b',c'}^{(2)} (\Omega_{WF,SF},t)   \rangle \times \\
	& \mathcal{D}_{a,b}^{(2)*} (\Omega_{D,WF}) \mathcal{D}_{c,0}^{(2)*} (\Omega_{SF,i})  \mathcal{D}_{a',b'}^{(2)} (\Omega_{D,WF}) \mathcal{D}_{c',0}^{(2)} (\Omega_{SF,j}).
	\end{aligned}
\label{eq:CorrFuncWobbPre}
\end{equation}
The Master equation that solves the conditional probability is given by:
\begin{equation}
\frac{\partial}{\partial t} p(\Omega, t) = -D_W L_W^2 p(\Omega, t),
\label{eq:SmoluchowskiEq}
\end{equation}
where $D_W$ is the diffusion coefficient for the wobble motion and $L_W^2$ is the angular momentum operator:
\begin{equation}
L_W^2 = - \frac{1}{\sin \theta} \frac{\partial}{\partial \theta} \left( \sin \theta \frac{\partial}{\partial \theta} \right) - \frac{1}{\sin^2 \theta}  \frac{\partial^2}{\partial \varphi^2},
\label{eq:WobbAngularMomentumOp}
\end{equation}
with reflecting boundary condition at $\theta = \beta_{c}$, with $\beta_{c}$ the cone semi-angle opening:
\begin{equation}
\frac{\partial}{\partial \theta} p(\Omega, t) |_{\theta=\beta_{c}} = 0.
\label{eq:BoundCondWob}
\end{equation}
\paragraph*{Solving the Master equation.} The eigen-equation is written:
\begin{equation}
D_W L_W^2 c(\theta, \varphi) = E c(\theta, \varphi),
\end{equation}
and we assume that the function $c$ can be separated into the product:
\begin{equation}
c(\theta, \varphi) = \Theta(\theta) \Phi(\varphi).
\end{equation}
This leads to the following differential equation for $\Phi$:
\begin{equation}
	\frac{\partial^2}{\partial \varphi^2} \Phi(\varphi) \propto \Phi(\varphi).
\end{equation}
Following the work of Wang and Pecora \cite{Wang_JCP_1980}, we write the complete set of functions $\Phi$ as:
\begin{equation}
\Phi_m(\varphi) = A_m \cos(m\varphi) + B_m \sin(m\varphi),
\end{equation}
where $m \in \ \mathbb{N}$, and $A_m$ and $B_m$ are coefficients associated to the eigenvalue $m$. It follows:
\begin{equation}
\frac{\partial^2}{\partial \varphi^2} \Phi_m(\varphi) = - m^2 \Phi_m(\varphi),
\end{equation}
and we can write the differential equation for $\Theta$ after the variable change $\mu \rightarrow \cos \theta$:
\begin{equation}
  2 \mu  \frac{\partial \Theta}{\partial \mu} - (1 - \mu^2) \frac{\partial^2 \Theta}{\partial \mu^2}  + \frac{m^2}{1 - \mu^2} \Theta - \frac{E}{D_W} \Theta = 0.
\end{equation}
We now define $E = D_W \nu_m (\nu_m + 1)$ to obtain:
\begin{equation}
 (1 - \mu^2) \frac{\partial^2 \Theta}{\partial \mu^2} - 2 \mu  \frac{\partial \Theta}{\partial \mu}  +\left( \nu_{m}(\nu_m + 1) -  \frac{m^2}{1 - \mu^2}  \right)\Theta = 0.
\label{eq:DiffEqWobbling}
\end{equation}
Solutions of the equations of the type Eq.\,\ref{eq:DiffEqWobbling} for $\mu \in [0, 1]$ are given by the Legendre associated functions $P_{\nu_m}^m$ of degree $\nu_m$ and order $m$. There are an infinite number of solutions and the eigenfunctions and eigenvalues are given by $P_{\nu_{m,n}}^m$ and $D_W \nu_{m,n} (\nu_{m,n} + 1)$ respectively, where the $\nu_{m,n}$ fulfill the boundary condition Eq.\,\ref{eq:BoundCondWob}. We choose the indices $n$ such that $\nu_{m,n} < \nu_{m,n+1}$. It can be noted that $\nu_{0,0} = 0$. The conditional probability $p(\Omega, t | \Omega_0, 0)$ is now given by:
\begin{equation}
p(\Omega, t | \Omega_0, 0) = \sum_{m=0}^{+\infty} \sum_{n = 0}^{+\infty} e^{-D_W \nu_{m,n}(\nu_{m,n} + 1) t}P_{\nu_{m,n}}^m (\cos \theta) (A_{m,n} \cos(m\varphi) + B_{m,n} \sin(m\varphi)) ,
\label{eq:WobCondProb}
\end{equation}
where we explicitely show the dependence in $m$ and $n$ of the coefficients $A$ and $B$. 
\paragraph*{Legendre associated functions.}
We give here the analytical expression of Legendre associated functions for $m \geq 0$:
\begin{equation}
P_{\nu_{m,n}}^m(z) = (-1)^{m} \frac{\Gamma(\nu_{m,n} + m + 1 )}{\Gamma(\nu_{m,n} - m + 1)} \frac{(1 - z)^{m/2}}{2^m m!}  _2{F}{_1} (-\nu_{m,n} + m, \nu_{m,n}  + m + 1, m + 1,  \frac{1 - z}{2}),
\label{eq:DefinitionLegendreFunctions}
\end{equation} 
where $\Gamma$ is the Gamma function defined for all $z \neq  0$:
\begin{equation}
\Gamma(z) = \int_0^\infty e^{-t}t^{z-1} dt,
\end{equation}
and $_2{F}{_1} $ is the Hypergeometric function:
\begin{equation}
_2{F}{_1}  (\alpha, \beta, \gamma, z) = 1 + \frac{\alpha \beta}{\gamma} \frac{z}{1!} +   \frac{\alpha(\alpha + 1) \beta (\beta + 1)}{\gamma (\gamma +1 )} \frac{z^2}{2!} + \cdot \cdot \cdot.
\end{equation}
The Hypergeometric function can be written using the Gamma function in a compact form as:
\begin{equation}
	\begin{dcases}
_2{F}{_1}  (\alpha, \beta, \gamma, z) = 1,\,\,\,\alpha=0\,\,\mathrm{or}\,\,\beta=0, \\
_2{F}{_1}  (\alpha, \beta, \gamma, z) = \sum_{k=0}^\infty \frac{\Gamma(\alpha + k)\Gamma(\beta+ k) \Gamma(\gamma)}{\Gamma(\alpha) \Gamma(\beta) \Gamma(\gamma + k)} \frac{z^k}{k!}, \alpha \neq 0\,\,\mathrm{and}\,\,\beta \neq 0.
	\end{dcases}
\label{eq:DefinitionHypergeometric}
\end{equation}
The definition of legendre associated function can be extended to orders having negative values:
\begin{equation}
P_{\nu_{-m,n}}^{-m}(z) = (-1)^m \frac{\Gamma(\nu_{m,n} - m + 1)}{\Gamma(\nu_{m, n} + m + 1)}P_{\nu_{m,n}}^m(z), m \geq 0.
\end{equation}
\paragraph*{Solving the boundary condition.} 
The conditional probability at time $t=0$ is:
\begin{equation}
	\begin{aligned}
p(\Omega, t=0 | \Omega_0, 0)  =& \delta(\mu - \mu_0) \delta(\varphi - \varphi_0), \\
	 = &\sum_{m=0}^{+\infty} \sum_{n = 0}^{+\infty}P_{\nu_{m,n}}^m (\mu_0)  (A_{m,n} \cos(m\varphi_0) + B_{m,n} \sin(m\varphi_0)),
	\end{aligned}
\end{equation}
where we set $\mu = \cos \theta$. Now, we can calculate the following two integrals:
\begin{equation}
	\begin{aligned}
	&I_c = \int_{\mu_{c}}^1 d\mu \int_0^{2\pi} d \varphi p(\Omega, t=0 | \Omega_0, 0) P_{\nu_{m',n'}}^{m'} (\mu) \cos m'\varphi, \\
	& I_s = \int_{\mu_{c}}^1 d\mu \int_0^{2\pi} d \varphi  p(\Omega, t=0 | \Omega_0, 0) P_{\nu_{m',n'}}^{m'} (\mu) \sin m'\varphi.
	\end{aligned}
\end{equation}
On one side, we have:
\begin{equation}
	\begin{aligned}
	& I_c = P_{\nu_{m',n'}}^{m'} (\mu_0) \cos m'\varphi_0, \\
	& I_s = P_{\nu_{m',n'}}^{m'} (\mu_0) \sin m'\varphi_0.
	\end{aligned}
\label{eq:ProbCoeffWobb}
\end{equation}
On the other side, we can write:
\begin{equation}
	\begin{aligned}
	I_c = H_{m',n'}(\mu_{c}) &\int_0^{2\pi} d \varphi   \cos m'\varphi  (A_{m',n'} \cos(m\varphi)  + B_{m',n'} \sin(m'\varphi)), \\
	I_s = H_{m',n'}(\mu_{c}) & \int_0^{2\pi} d \varphi  \sin m'\varphi   (A_{m',n'} \cos(m\varphi) + B_{m',n'} \sin(m'\varphi)),
	\end{aligned}
\end{equation}
where $\mu_c = \cos \beta_c$ and $H_{m,n}$ is defined from the orthogonality condition of Legendre associated functions \cite{Wang_JCP_1980}:
\begin{equation}
\int_{\mu_{c}}^1 P_{\nu_{m,n}}^m (\mu) P_{\nu_{m',n'}}^{m'} (\mu) d\mu = \delta_{nn'} \delta_{mm'} H_{m,n} (\mu_{c}).
\label{eq:LegendreOrthonormality}
\end{equation}
After integration and identification, we can express the two coefficients $A_{m,n}$ and $B_{m,n}$:
\begin{equation}
	\begin{aligned}
	& A_{m,n} = \frac{\cos m\varphi_0}{\epsilon \pi H_{m,n}(\mu_{c})} P_{\nu_{m,n}}^m (\mu_0), \\
	& B_{m,n} = \frac{\sin m\varphi_0}{ \pi H_{m,n}(\mu_{c})} P_{\nu_{m,n}}^m (\mu_0),
	\end{aligned}
\end{equation}
where $\mu_0 = \cos \theta_0$ and:
\begin{equation}
\epsilon = 
\begin{dcases}
2,\,\,\,m=0 \\
1,\,\,\,m=\{1,2\}
\end{dcases}.
\end{equation}
\paragraph*{Conditional probability.} The conditional probability is now written:
\begin{equation}
p(\Omega, t | \Omega_0, 0) = \sum_{m=0}^{+\infty} \sum_{n = 0}^{+\infty}  X_{m,n},
\label{eq:CondProbWobbInit}
\end{equation}
where we define:
\begin{equation}
X_{m,n} = e^{-D_W \nu_{m,n}(\nu_{m,n} + 1) t} \frac{P_{\nu_{m,n}}^m (\mu) P_{\nu_{m,n}}^m (\mu_0)}{\pi H_{m,n}(\mu_{c})}   \left(\frac{\cos m\varphi_0}{\epsilon} \cos m\varphi + \sin m\varphi_0 \sin m\varphi \right).
\end{equation}
We have:
\begin{equation}
X_{0,n} = \frac{1}{2}  \frac{P_{\nu_{0,n}}^0 (\mu) P_{\nu_{0,n}}^0 (\mu_0)}{\pi H_{0,n}(\mu_{c})} e^{-D_W \nu_{0,n}(\nu_{0,n} + 1) t}
\label{eq:DefX0n}
\end{equation}
and, for $m \neq 0$:
\begin{equation}
X_{m,n} = e^{-D_W \nu_{m,n}(\nu_{m,n} + 1) t} \frac{P_{\nu_{m,n}}^m (\mu) P_{\nu_{m,n}}^m (\mu_0)}{\pi H_{m,n}(\mu_{c})}  \left(\cos m\varphi_0\cos m\varphi + \sin m\varphi_0 \sin m\varphi\right).
\label{eq:DefXmn}
\end{equation}
We use the following properties for Legendre associated functions and their roots:
\begin{equation}
	\begin{aligned}
		&\nu_{-m,n} = \nu_{m,n}, \\
		&\frac{P_{\nu_{-m,n}}^{-m}(\mu_1) P_{\nu_{-m,n}}^{-m}(\mu_2)}{H_{-m,n}(\mu_{c})} = \frac{P_{\nu_{m,n}}^{m}(\mu_1) P_{\nu_{m,n}}^{m}(\mu_2)}{H_{m,n}(\mu_{c})},
	\end{aligned}
\end{equation}
to write:
\begin{equation}
p(\Omega, t | \Omega_0, 0) = \frac{1}{2} \sum_{m=-\infty}^{+\infty} \sum_{n = 0}^{+\infty} e^{-D_W \nu_{m,n}(\nu_{m,n} + 1) t}  \frac{P_{\nu_{m,n}}^m (\mu) P_{\nu_{m,n}}^m (\mu_0)}{\pi H_{m,n}(\mu_{c})} e^{i m \varphi} e^{-i m \varphi_0}.
\end{equation}
The factor $1/2$ comes from the equality $X_{-m,n} = X_{m,n}$ and the extension of the sum on the index $m$ from $-\infty$ to $+\infty$, and the expression of $X_{0,n}$ (Eq.\,\ref{eq:DefX0n}). C.\,Wang and R.\,Pecora introduced the pseudo-spherical harmonics \cite{Wang_JCP_1980}:
\begin{equation}
Y_{\nu_{m,n}}^m (\Omega) = \frac{1}{\sqrt{2\pi H_{m,n}(\mu_{c})}} P_{\nu_{m,n}}^m (\mu) e^{i m \varphi},
\end{equation}
and wrote the conditional probability in a compact form:
\begin{equation}
p(\Omega, t | \Omega_0, 0) = \sum_{m=-\infty}^{+\infty} \sum_{n = 0}^{+\infty} e^{-D_W \nu_{m,n}(\nu_{m,n} + 1) t}   Y_{\nu_{m,n}}^m (\Omega) Y_{\nu_{m,n}}^{m*} (\Omega_0).
\end{equation}
\paragraph*{Initial angle probability.}
The conditional probability does not cancel out when $t \rightarrow \infty$ only for $\nu_{m,n}=0$, that is $m=n=0$:
\begin{equation}
p(\Omega_0) = Y_0^0 (\Omega) Y_0^0 (\Omega_0) = \frac{P_0^0 (\mu) P_0^0 (\mu_0)}{2 \pi H_{0,0}(\mu_{c})}.
\end{equation}
It is then straightforward to show:
\begin{equation}
	\begin{dcases}
	p(\Omega_0) = \frac{1}{2\pi (1 - \mu_{c})},\,\,\,\theta_0 \in [0, \beta_{c}], \\
	p(\Omega_0) = 0,\,\,\, \theta_0 > \beta_{c}.
	\end{dcases}
\end{equation}
The correlation function for wobbling in a cone then reads:
\begin{equation}
	\begin{aligned}
& \langle \mathcal{D}_{b,c}^{(2)*} (\Omega_{WF,SF},0) \mathcal{D}_{b',c'}^{(2)} (\Omega_{WF,SF},t)   \rangle = \frac{1}{4\pi^2 (1-\mu_{c})}  \sum_{m=-\infty}^{+\infty} \sum_{n=0}^{+\infty} \frac{e^{-D_W \nu_{m,n} (\nu_{m,n} + 1) t}}{H_{m,n}(\mu_{c})} \int_0^{2\pi}  d\varphi_0  \int_0^{2\pi}  d\varphi \times \\
	& \int_0^{\beta_{c}} \sin \theta_0 d\theta_0 \int_0^{\beta_{c}} \sin \theta d\theta P_{\nu_{m,n}}^m (\cos \theta_0) P_{\nu_{m,n}}^m (\cos \theta)   \mathcal{D}_{b,c}^{(2)*} (\varphi_0, \theta_0, -\varphi_0) \mathcal{D}_{b',c'}^{(2)} (\varphi, \theta, -\varphi) e^{i m (\varphi - \varphi_0)}.
	\end{aligned}
\end{equation}
The third Euler angles in the Wigner matrices equal the opposite of the first Euler angles in order for this correlation function to only account for the motion of the C-C bond in a cone, and not any rotational motions around the z-axis of the SF: if the third angle was set to 0, the orientation of the x-axis of the system frame with respect to the wobbling frame main axis would not change upon diffusion in the cone thus introducing an additional rotation motion. When it was initially developped, this model was accounting for motions of the interaction of interest \cite{Kinosita_BiophysJ_1977,Warchol_AdvMolRelax_1978}. In this situation, the second index of the Wigner matrices equals 0, and the third Euler angle does not play any role. This is not the case when the interaction frame is not the one undergoing the wobbling motion, and introducing this angle is essential to best describe the motions.\\
Expanding the expression  of the Wigner matrices and performing the integration on $\varphi_0$ and $\varphi$ leads to the condition $m=b-c=b'-c'$. Note that in the previous models, the condition $m=b=b'$ was obtained, which is correct when the interaction frame is the diffusing frame since $c=c'=0$ \cite{Wang_JCP_1980,Kumar_JCP_1986,Kumar_JCP_1989}. We finally can write the total correlation function as:
\begin{equation}
	\begin{aligned}
C_{i,j}(t) = \frac{1}{5} &\sum_{\kappa=1}^5 \sum_{a,a',b,b',c,c'=-2}^2 \delta_{b-c, b'-c'}  a_{\kappa,a} a_{\kappa,a'} e^{-E_\kappa t}  \sum_n \frac{e^{-D_W \nu_{b-c,n} (\nu_{b-c,n} + 1) t}}{(1 - \mu_{c})H_{b-c,n}(\mu_{c})}  I_{b,c}^n (\beta_{c}) I_{b',c'}^n (\beta_{c}) \times  \\
& \mathcal{D}_{a,b}^{(2)*} (\Omega_{D,WF}) \mathcal{D}_{c,0}^{(2)*} (\Omega_{SF,i})  \mathcal{D}_{a',b'}^{(2)} (\Omega_{D,WF}) \mathcal{D}_{c',0}^{(2)} (\Omega_{SF,j}) ,
	\end{aligned}
\end{equation}
where:
\begin{equation}
I_{b,c}^n(\beta_{c}) = \int_0^{\beta_{c}}  \sin \theta d_{b,c}(\theta)  P_{\nu_{b-c,n}}^{b-c} (\cos \theta)  d\theta.
\label{eq:IbcnDef}
\end{equation}
\paragraph*{Order parameter for wobbling in a cone.}
The correlation function for wobbling is non-zero when $t \rightarrow +\infty$ for $b-c = n = 0$. Thus, we have:
\begin{equation}
\mathcal{S}_W^2 (i, j) = \sum_{b=-2}^2 \left(\frac{I_{b,b}^0(\beta_{c})}{1 - \mu_{c}} \right)^2  \mathcal{D}_{b,0}^{(2)*} (\Omega_{SF,i})\mathcal{D}_{b,0}^{(2)} (\Omega_{SF,j}).
\label{eq:WobbOrderParamFull}
\end{equation}
In the particular case of auto-correlation and where the PAF of the interactions are undergoing the wobbling motions, we have $\Omega_{SF,i} = \Omega_{SF,j} = \{0,0,0\}$ and the order parameter simplifies into the equation reported by K.\,Kinosita \textit{et al.} \cite{Kinosita_BiophysJ_1977}:
\begin{equation}
\mathcal{S}_W^2 (i_W, i_W) = \frac{1}{4} \cos^2 \beta_{c} (1 + \cos \beta_{c})^2,
\label{eq:KinositaOrderParam}
\end{equation}
where we write $i_W$ to indicate that the interaction $i$ is diffusing in a cone.

\subsection{Contribution of each motion to relaxation}
\begin{figure*}[!ht]
		\includegraphics[width=0.95\textwidth]{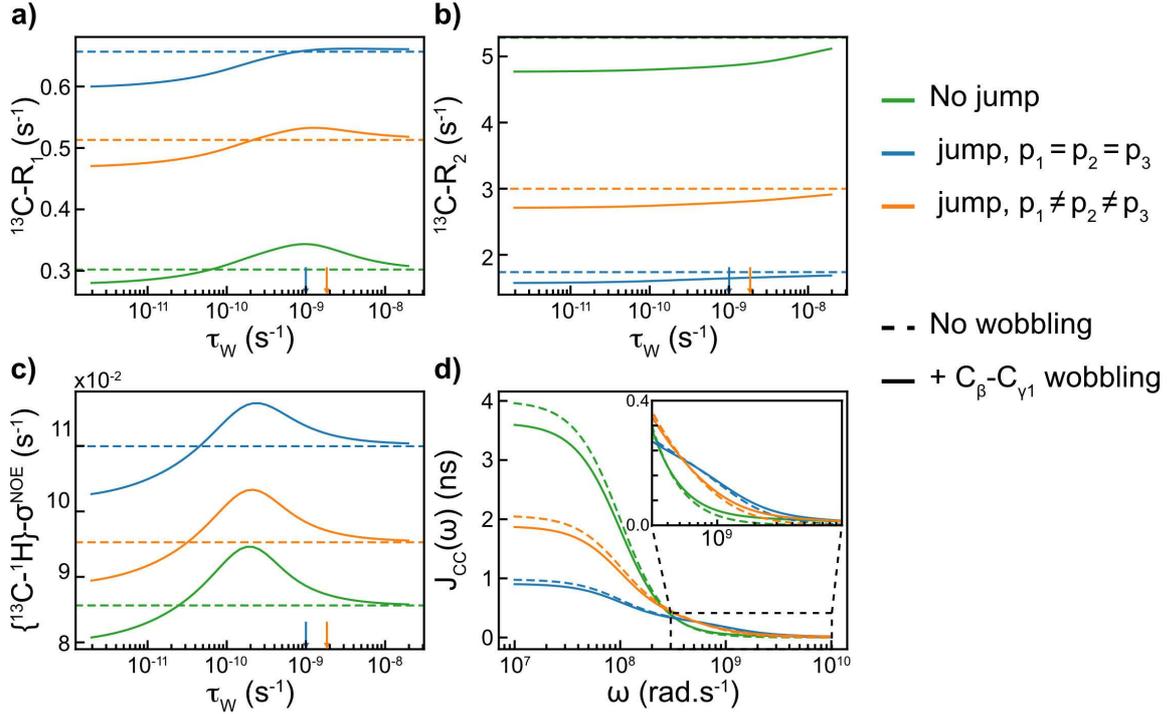}
	\caption{Contribution of rotamer jump and wobbling in a cone motions to relaxation. Carbon R$_1$ (\textbf{a}), R$_2$ (\textbf{b}) and carbon-proton $\sigma^\mathrm{NOE}$ (\textbf{c}) at 14.1\,T in a hypothetical $^{13}$C$^1$H$^2$H$_2$ valine methyl group, and associated spectral density function for the C$_{\beta}$-C$_{\gamma1}$ bond auto-correlations. Dashed lines show the value of the relaxation rates and spectral density function without wobbling. Relaxation rates (\textbf{a}-\textbf{c}) and spectral density function (\textbf{d}) calculated in the abscence of rotamer jump are shown in green. In the presence of rotamer jump, the case where all rotamer populations are equal (blue) and unequal (orange) are distinguished. In this latter case, populations are $p_1=0.7$, $p_2=0.2$ and $p_1=0.1$. Calculations are shown as a function of the correlation time for wobbling, a function of the wobbling diffusion constant and cone semi-angle opening, set here to $\beta_{c}=15$\,deg, while the diffusion constant $D_W$ is varied from $10^6$ to $10^{10}$\,s$^{-1}$. The blue and orange vertical arrows indicate the values of the correlation time for rotamer jump, respectively when populations are equal and unequal. The spectral density functions are shown for $D_W = 10^8$\,s$^{-1}$.}
	\label{fig:ContributionMotions}
\end{figure*}
We evaluate here how each motion contributes to relaxation. Having an exhaustive analysis is not feasible and we will focus on typical relaxation rates: $^{13}$C-R$_1$,  $^{13}$C-R$_2$ and $^{13}$C-$^1$H-$\sigma^\mathrm{NOE}$ in a $^{13}$C$^1$H$^2$H$_2$ methyl group as encountered in a valine side-chain. The overall diffusion is supposed to be isotropic with global tumbling correlation time $\tau_c = 10$\,ns. The methyl rotation diffusion coefficient is set to $D_{rot} = 5\times10^{10}$\,s$^{-1}$, which leads to a correlation time for rotation of $\tau_{rot}=10$\,ps. When a jump-type of motion is added, we suppose that the valine side-chain can jump between three possible rotamer states, and we distinguish the situation where each rotamers are equaly populated with equal exchange rate $k_{ex}^{sym}=1/3\times10^9$\,s$^{-1}$ between the different conformers (which leads to a correlation time for exchange of $\tau_{ex}^{sym}=1$\,ns), from the situation where $p_1=0.7$, $p_2=0.2$ and $p_3=0.1$ (the state numerotation is irrelevant since the global diffusion tensor is isotropic) and all exchange rates $k_{ij}^{asym}=1/3\times10^9$\,s$^{-1}$ with $i < j$ and $k_{ij}^{asym}$ with $i>j$ is calculated to satisfy the microscropy reversibility condition (Eq.\,\ref{eq:MicroRev}) (which leads to a correlation time for exchange of $\tau_{ex}^{asym}=1.8$\,ns). The Euler angles associated to the jump motion are $\Omega_J = \{2n\pi/3, \beta_{CC}, 0\}$ where $n=0,1,2$ and $\beta_{CC} = 76$\,deg, which is typical for carbon side-chains. A wobbling in a cone motion of the C$_{\beta}$-C$_{\gamma1}$ bond is also considered, when mentioned. The cone semi-angle opening is set to 15\,deg (results for opening angles of 5, 30 and 60\,deg are shown in Fig.\,S3), and the associated diffusion constant $D_W$ is varied from $10^6$ to $10^{10}$\,s$^{-1}$, leading to variations of the wobbling correlation time from 1\,ps to 10\,ns approximately for this cone angle opening. Relaxation rates are calculated at a magnetic field of 14.1\,T, and the carbon-13 CSA is set to 20\,ppm \cite{Tugarinov_JBNMR_2004}. \\
The rotamer exchange contributes significantly to the relaxation (compare dashed lines in Fig.\,\ref{fig:ContributionMotions}), with an R$_1$ and $\sigma^\mathrm{NOE}$ increase by a factor 1.5 approximately, while R$_2$ decrease by a factor up to 3.8 when rotamer jumps with equal populations are introduced. The effect of rotamer jumps is reduced for R$_2$ and $\sigma^\mathrm{NOE}$ when the populations are not equal, and increased for R$_1$. These effects can be rationalized by inspecting the spectral density functions (Fig.\,\ref{fig:ContributionMotions}d). In the presence of internal motions, the spectral density function tends to be higher at frequencies in the range $10^9$-$10^{10}$\,rad.s$^{-1}$, which are frequencies R$_1$ and $\sigma^\mathrm{NOE}$ are sensitive to, leading to an increase of these two rates. The higher contribution of medium-frequency motions to the spectral density function goes in pair with a decrease in the contribution of low-frequency motions, which is the main determinant in the value of the transverse relaxation rate $R_2$, explaining the large effects calculated in the presence of rotamer exchange. \\
On the other hand, the wobbling of the C$_{\beta}$-C$_{\gamma1}$ bond has smaller effects on relaxation (Fig.\,\ref{fig:ContributionMotions}) for the choosen opening of the cone. Wobbling reduces the transverse relaxation rate $R_2$, but the effect on the longitudinal auto- and cross-relaxation rates depends on the value of the diffusion constant $D_W$: fast diffusion (low $\tau_W$) leads to a decrease in both R$_1$ and $\sigma^\mathrm{NOE}$; we obtain an increase for lower values of $D_W$ (higher values of $\tau_W$); and hardly any effects for very slow diffusion. These results can be rationalized using the same arguments as for the rotamer jump by analyzing the evolution of the spectral density function and the relative contribution of each frequency range to the value of relaxation rates. It is interesting to notice the similarities between the evolution of the longitudinal relaxation rate R$_1$ and $\sigma^\mathrm{NOE}$ as a function of $\tau_W$ and the sensitivities calculated in the detector approach which characterizes the amount of motions in a given range of correlation times \cite{Smith_JCP_2018,Smith_JCP_2019,Smith_anie_2019,Smith_JBNMR_2021}. \\
The contribution of the wobbling motions increases with increasing cone semi-angle opening and can even become larger than the contribution of the rotamer jump (Fig.\,S3). This can be explained by the value of order parameters, which act as weights for the Lorentzian terms of the spectral density containing contribution only from the global tumbling. The order parameters for rotamer exchange when populations are equal and unequal are 0.17 and 0.43 respectively. In comparison, the wobbling order parameter equals 0.99, 0.90, 0.65 and 0.14 for values of cone semi-angle opening of 5, 15, 30 and 60\,deg. Thus, for $\beta_{c}=60$\,deg, the decrease in $R_2$ from either a  rotamer exchange with equal population or the wobbling are similar for $\tau_W \approx \tau_{ex}^{sym}$ (Fig.\,S3.h). The conclusions also apply when an alternative model of motion is used where the wobbling of the C$_\alpha$-C$_\beta$ is allowed but the C$_\beta$-C$_\gamma$ is fixed in the rotamer frame (Fig.\,S4). These two models may be difficult to discrimitate with experimental data. \\
As a general rule of thumb, one can reasonably expect rotamer jumps to be the major source of relaxation for methyl groups in aliphatic side-chains, since the amplitudes of wobbling are usually limited to small cone angle openings. This has already been noted by Wand and coworker \cite{Frederick_JCPB_2008}, but of course has to be carefully investigated before neglecting one motion, for example with the use of MD simulations.

%% file: Section/4_CorrelatedMotions.tex
\section{Correlated motions}
So far, correlation functions were written assuming statistical independence between each motion. This assumption does not hold in some cases. Correlation functions for correlated overal tumbling and jumps of entire domain have been presented in the late 2000s \cite{Wong_PNAS_2009, Ryabov_JCP_2012,Gill_JPCB_2014,Ryabov_Proteins_2015}. In the case of methyl groups in aliphatic side-chains, the methyl rotation or C-C wobbling can potentially depend on the rotameric state, introducing correlation of \textit{internal} motions \cite{Straus_JBNMR_1997,BolikCoulon_ArXiv_2022,Smith_JMRO_2022}. We show here how to write the correlation function in the presence of such correlated motions.

\subsection{Correlated jumps and methyl rotation}
\begin{figure*}[!ht]
	\begin{center}
		\includegraphics[width=0.9\textwidth]{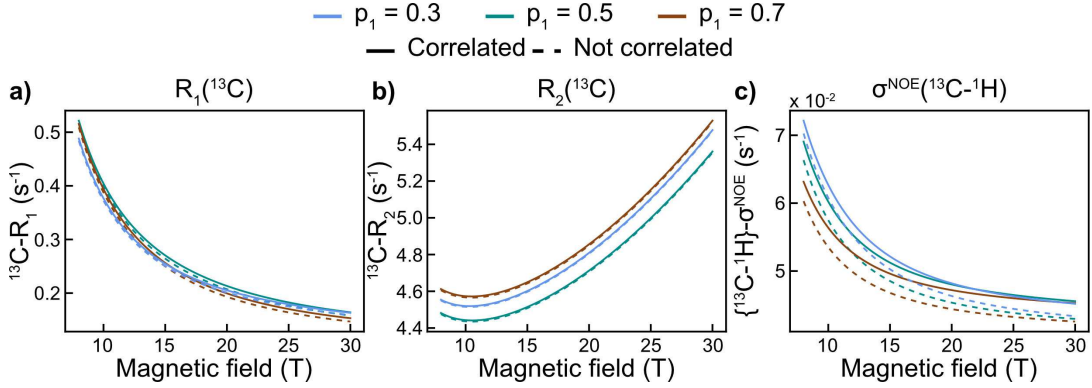}
	\end{center}
	\caption{Contribution of correlated rotamer jumps and methyl rotation to relaxation. Magnetic field variations of carbon-13 R$_1$ (\textbf{a}), R$_2$ (\textbf{b}) and carbon-13-proton DD cross-relaxation rate (\textbf{c}) in a $^{13}$C$^1$H$^2$H$_2$ methyl group exchanging between two rotamers states and calculated for three distributions of populations. Relaxation rates calculated with a unique diffusion coefficient for methyl rotation are shown in dash, and relaxation rates calculated by considering the correlation between rotamer jumps and methyl rotation motions are shown in plain lines. The overall diffusion is isotropic with global tumbling correlation time $\tau_c = 10$\,ns. The population average methyl rotation coefficient is $10^{11}$\,s$^{-1}$ and the difference between the two state is $5 \times 10^{10}$\,s$^{-1}$. The rotamer 1 has the highest diffusion coefficient. Carbon-13 CSA is set to 20\,ppm.}
	\label{fig:CorrelatedRotJump}
\end{figure*}
Using the same notations as above,  the correlation funtion for correlated rotamer jumps and methyl rotation is:
\begin{equation}
	\begin{aligned}
C_{i,j}(t) = \frac{1}{5} &\sum_{\kappa=1}^5 \sum_{a,a',b,b',c,c'=-2}^2 a_{\kappa,a} a_{\kappa,a'} e^{-E_\kappa t}\langle \mathcal{D}_{b,c}^{(2)*} (\Omega_{J,SF},0) \mathcal{D}_{b',c'}^{(2)} (\Omega_{J,SF}, t) \rangle \times  \\
	 &\mathcal{D}_{a,b}^{(2)*} (\Omega_{D,J}) \mathcal{D}_{a',b'}^{(2)} (\Omega_{D,J}) \mathcal{D}_{c,0}^{(2)*} (\Omega_{SF,i})  \mathcal{D}_{c',0}^{(2)} (\Omega_{SF,j})  .
	\end{aligned}
\label{eq:CorrMotionsGenEq}
\end{equation}
In the Euler angle set $\Omega_{J,SF}=\{\varphi_{J,SF}, \theta_{J,SF}, \phi_{J,SF}\}$, $\varphi_{J,SF}$ rotates the jump frame to the direction of the populated rotamer and is \textit{time dependent}, $\theta_{J,SF}$ rotates the jump frame to align it on the rotamer frame, and $\phi_{J,SF}$ rotates the resulting frame to align its x-axis along the direction of one C-H bond and is \textit{time dependent}. We assume that both interactions $i$ and $j$ undergo rotation such that $\Omega_{SF,i} = \Omega_{SF,j}$. Note that in the case where only one interaction undergoes rotation, as in methyl-groups CSA/DD cross-correlated cross-relaxation, the resulting correlation function is independent from the rotation (Eq.\,\ref{eq:CSADDcrosscorrelation}) such that there is no possible correlation between jumps and rotation.  The ensemble average can be expressed as:
\begin{equation}
 \langle \mathcal{D}_{b,c}^{(2)*} (\Omega_{J,SF},0) \mathcal{D}_{b',c'}^{(2)} (\Omega_{J,SF}, t) \rangle  = \sum_{\alpha, \beta=1}^N p_\alpha^{eq}  \langle \mathcal{D}_{b,c}^{(2)*} (\Omega_{J,SF},0) \mathcal{D}_{b',c'}^{(2)} (\Omega_{J,SF}, t) \rangle_{\beta \alpha} ,
\label{eq:CorrMotionsCompleteEnsembleSum}
\end{equation}
where the sums run over all accessible states, and the notation $\langle \cdot \cdot \cdot \rangle_{\beta \alpha}$ indicates that the ensemble average is calculated with initial state $\alpha$ and final state $\beta$. It can be calculated as follows:
\begin{equation}
	\begin{aligned}
 \langle \cdot \cdot \cdot \rangle_{\beta \alpha} = & \int_0^{2\pi} d \phi_0 \int_0^{2\pi} d \phi   p(\phi_0)  p(\{ \varphi_{J,SF_\beta}, \theta_{J,SF_\beta}, \phi \}, t | \{\varphi_{J,SF_\alpha}, \theta_{J,SF_\alpha}, \phi_0\}, 0) \times  \\
	& \mathcal{D}_{b,c}^{(2)*} (\{\varphi_{J,SF_\alpha}, \theta_{J,SF_\alpha}, \phi_0\})  \mathcal{D}_{b',c'}^{(2)} (\{\varphi_{J,SF_\beta}, \theta_{J,SF_\beta}, \phi\}), 
	\end{aligned}
\label{eq:EnsembleAverageCorrRotEx}
\end{equation}
where $\varphi_{J,SF_k}$ and $\theta_{J,SF_k}$ are the values of the Euler angles to transform from the jump frame to the system frame in rotamer $k$. The probability $p(\phi_0)$ is:
\begin{equation}
p(\phi_0) = \frac{1}{2\pi},
\end{equation}
and in order to calculate the conditional probability, we will use the notation:
\begin{equation}
 p(\{ \varphi_{J,SF_\beta}, \theta_{J,SF_\beta}, \phi \}, t | \{\varphi_{J,SF_\alpha}, \theta_{J,SF_\alpha}, \phi_0\}, 0) = 
 p(\beta, \phi, t | \alpha, \phi_0, 0).
\end{equation}
The Master equation that solves the conditional probability is a combination of the Master equations for rotamer exchange and diffusion on a cone:
\begin{equation}
 \frac{d}{dt} p(\beta, \phi, t | \alpha, \phi_0, 0) = \sum_{\gamma=1}^N R_{\beta \gamma} p(\gamma
, \phi, t | \alpha, \phi_0, 0) - D_{rot,\beta} L_{rot}^2 p(\beta, \phi, t | \alpha, \phi_0, 0),
\label{eq:MasterEqCorrRotEx}
\end{equation}
where $R_{\beta\gamma}$ are elements of the exchange matrix $\mathcal{R}$ (\textit{i.e.} exchange rate from state $\gamma$ to state $\beta$) and $D_{rot,\beta}$ is the diffusion constant for diffusion on a cone in state $\beta$. We solve Eq.\,\ref{eq:MasterEqCorrRotEx} by writing the conditional probability in terms of eigenfunctions of the angular momentum operator $L_{rot}^2$:
\begin{equation}
p(\beta, \phi, t | \alpha, \phi_0, 0) = \sum_n \frac{1}{2\pi} e^{in(\phi - \phi_0)} c_n^{\beta \alpha} (t),
\end{equation}
which, after insertion and identification in Eq.\,\ref{eq:MasterEqCorrRotEx}, leads to the following differential equation for the functions $c_n^{\beta \alpha}$:
\begin{equation}
\frac{d}{dt} c_n^{\beta \alpha} (t) = - D_{rot, \beta} n^2 c_n^{\beta \alpha} (t) + \sum_{\gamma = 1}^N R_{\beta \gamma} c_n^{\gamma \alpha}(t).
\end{equation}
It can be written in matrix form as:
\begin{equation}
\frac{d}{dt} C_n^\alpha (t) = \left( \mathcal{R} - n^2 \mathcal{D}_{rot}\right) C_n^\alpha (t),
\end{equation}
where $\mathcal{D}_{rot}$ is a diagonal matrix containing the methyl rotation diffusion coefficients as diagonal elements and $C_n^\alpha (t)$ is a column vector containing the elements $c_n^{\beta \alpha}(t)$ for all states $\beta$. Similarly to the treatment of rotamer jumps, we define the symmetrized pseudo-exchange matrix as:
\begin{equation}
\tilde{\mathcal{R}}_{rot,n} = \tilde{\mathcal{R}} - n^2 \mathcal{D}_{rot},
\end{equation}
where $\tilde{\mathcal{R}}$ is the symmetrized exchange matrix (Eq.\,\ref{eq:SymExMatrixDefinition}). Then, the functions $c_n^{\beta \alpha}$ can be explicitely written as:
\begin{equation}
c_n^{\beta \alpha} (t) = \sqrt{\frac{p_\beta^{eq}}{p_\alpha^{eq}}} \sum_{m=1}^N \tilde{X}_\alpha^{(n,m)} \tilde{X}_\beta^{(n,m)} e^{\lambda_{n,m} t},
\end{equation}
where $N$ is the number of states and $\tilde{X}_\zeta^{(m,n)}$ is the value of the eigenvector $\tilde{X}^{(n,m)}$ associated to the eigenvalue $\lambda_{n,m}$ and at coordinate $\zeta$. After integration, we obtain the correlation function for correlated rotamer jumps and methyl rotation:
\begin{equation}
	\begin{aligned}
C_{i,j}(t) = \frac{1}{5} \sum_{\kappa=1}^5 \sum_{a,a',b=-2}^2 \sum_{\alpha, \beta} \sum_n a_{\kappa,a} a_{\kappa,a'} e^{-E_\kappa t} \sqrt{p_\alpha^{eq}p_\beta^{eq}} \tilde{X}_\alpha^{(b,n)} \tilde{X}_\beta^{(b,n)} e^{\lambda_{b,n}t} \times & \\
		\mathcal{D}_{a,b}^{(2)*} (\Omega_{D,SF_\alpha}) \mathcal{D}_{a',b}^{(2)} (\Omega_{D,SF_\beta}) \left[ d_{b,0}(\beta_{SF,i}) \right]^2 .&
	\end{aligned}
\end{equation}
In order to investigate how the correlation of motions affects relaxation rates in a $^{13}$C$^1$H$^2$H$_2$ methyl group, we consider a simple 2-state exchange, with Euler angles for jumps of $\{\pm \pi/2, \beta_{J,CC},0\}$, where $\beta_{JCC}=76$\,deg is typical for aliphatic carbon chains and thus corresponds to a $2\beta_{JCC}$-jump. The exchange rate $k_{12}$ is fixed to $k_{12}=0.5\times10^9$\,s$^{-1}$, and $k_{21}$ is calculated using Eq.\,\ref{eq:MicroRev} to satisfy the miscroscopic reversibility condition. We impose an isotropic overall diffusion tensor with global tumbling correlation time $\tau_c=10$\,ns. Relaxation rates are calculated for three distributions of rotamers: $p_1=0.3$, $p_1=0.5$ and $p_1=0.7$. Finally, the population average methyl rotation diffusion coefficient is set to $10^{11}$\,s$^{-1}$, and the difference of diffusion coefficients between the two states is $\Delta D_{rot} = 5\times10^{10}$\,s$^{-1}$, with the highest diffusion coefficient being associated to rotamer 1. We impose an axially symmetric CSA tensor for the $^{13}$C nucleus with an anisotropy of 20\,ppm. \\
Transverse relaxation rates calculated with the same properties for methyl rotation in both rotamers are indistinguishable from the rates where the correlation between motions are taken into account (Fig.\,\ref{fig:CorrelatedRotJump}b). However, longitudinal auto-relaxation rates (Fig.\,\ref{fig:CorrelatedRotJump}a) and DD cross-relaxation rates (Fig.\,\ref{fig:CorrelatedRotJump}c) show significant deviations depending on whether correlated motions are considered or not. \\

\subsection{Correlated jumps and wobbling in a cone}
Here we treat the case where C-C wobbling depends on the rotamer state. The correlation function can still be written as in Eq.\,\ref{eq:CorrMotionsGenEq} but the angles $\varphi_{J,SF}$, $\theta_{J,SF}$ and $\phi_{J,SF}$ are all time-dependent, which complicates the subsequent integration as in Eq.\,\ref{eq:EnsembleAverageCorrRotEx}. In order to overcome this difficulty, we decompose the Wigner matrices in Eq.\,\ref{eq:CorrMotionsCompleteEnsembleSum} into:
\begin{equation}
	\begin{aligned}
& \langle \mathcal{D}_{b,d}^{(2)*} (\Omega_{J,SF},0) \mathcal{D}_{b',d'}^{(2)} (\Omega_{J,SF}, t) \rangle  =  \sum_{\alpha, \beta=1}^N p_\alpha^{eq} \times  \\
		&\sum_{c, c'}  \langle \mathcal{D}_{b,c}^{(2)*} (\Omega_{J,W},0) \mathcal{D}_{b',c'}^{(2)} (\Omega_{J,W}, t)  \mathcal{D}_{c,d}^{(2)*} (\Omega_{W,SF},0) \mathcal{D}_{c',d'}^{(2)} (\Omega_{W,SF}, t) \rangle_{\beta \alpha} ,
	\end{aligned}
\end{equation}
where the Wigner matrices with Euler angles $\Omega_{J,W}$ for transformation from the jump frame to the wobbling frame can effectively be taken out of the correlation function $\langle \cdot \cdot \cdot \rangle_{\beta \alpha}$ since they account for the jump from state $\alpha$ to $\beta$:
\begin{equation}
	\begin{aligned}
 \langle \mathcal{D}_{b,d}^{(2)*} (\Omega_{J,SF},0) \mathcal{D}_{b',d'}^{(2)} (\Omega_{J,SF}, t) \rangle  = & \sum_{\alpha, \beta=1}^N p_\alpha^{eq} \sum_{c, c'}  \mathcal{D}_{b,c}^{(2)*} (\Omega_{J,W_\alpha}) \mathcal{D}_{b',c'}^{(2)} (\Omega_{J,W_\beta})   \times  \\
& \langle \mathcal{D}_{c,d}^{(2)*} (\Omega_{W,SF},0) \mathcal{D}_{c',d'}^{(2)} (\Omega_{W,SF}, t) \rangle_{\beta \alpha} ,
	\end{aligned}
\end{equation}
where $W_\alpha$ refers to the wobbling frame in state $\alpha$. We follow the same approach introduced in the case of correlated jumps and rotation on a cone to write the conditional probability $p(\varphi, \theta, \beta, t | \varphi_0, \theta_0, \alpha, 0)$:
\begin{equation}
	p(\varphi, \theta, \beta, t | \varphi_0, \theta_0, \alpha, 0) = \sum_{m=-\infty}^{+\infty} \sum_{n = 0}^{+\infty} \frac{e^{im (\varphi - \varphi_0)} \mathcal{P}_{\nu_{m,n}^\alpha}^m (\mu_0) \mathcal{P}_{\nu_{m,n}^\beta}^m (\mu)}{2\pi \sqrt{H_{m,n}(\mu_c^\alpha) H_{m,n}(\mu_c^\beta)}}  c_{m,n}^{\beta \alpha}(t), 
\end{equation}
where the Euler angle for transformation from the Wobbling frame (W) to the System Frame (SF) is written $\Omega_{W,SF} = \{\varphi, \theta, -\varphi\}$, $\mu = \cos \theta$ and $\mu_c^\alpha = \cos \beta_c^\alpha$ with $\beta_c^\alpha$ the cone semi-angle opening in state $\alpha$. The index $0$ refers to angles evaluated at time $t = 0$. The time-dependent function $c_{m,n}^{\beta\alpha}$ is found by solving:
\begin{equation}
\frac{d}{dt} C_{m,n}^\alpha (t) = (\mathcal{R} - \mathcal{D}_W^{(m,n)}) C_{m,n}^\alpha (t),
\end{equation}
where $C_{m,n}^\alpha (t)$ is a column vector containing the elements $c_{m,n}^{\beta \alpha}(t)$ for all states $\beta$, $\mathcal{R}$ is the exchange matrix and $\mathcal{D}_W^{(m,n)}$ is a diagonal matrix with diagonal elements defined as $D_{W,\beta} \nu_{m,n}^\beta (\nu_{m,n}^\beta + 1)$ for all states $\beta$. Using the same treatment as shown above, we obtain:
\begin{equation}
c_{m,n}^{\beta \alpha}(t) = \sqrt{\frac{p_\beta^{eq}}{p_\alpha^{eq}}} \sum_{l = 1}^N \tilde{X}_\alpha^{(n,m,l)} \tilde{X}_\beta^{(n,m,l)} e^{\lambda_{n,m,l} t},
\end{equation}
where the sum runs over all states $N$ and $\tilde{X}_\alpha^{(m,n,l)}$ is the $\alpha^\mathrm{th}$ value of the eigenvector $\tilde{X}^{(n,m,l)}$ associated to the eigenvalue $\lambda_{n,m,l}$ for the symmetrized pseudo-exchange matrix $\tilde{\mathcal{R}}_{W,m,n} = \tilde{\mathcal{R}} - \mathcal{D}_W^{(m,n)}$. Finally, the total correlation function can be calculated, and we obtain, using the same notations as above:
\begin{equation}
	\begin{aligned}
C_{i,j}(t) = \frac{1}{5} \sum_\kappa \sum_{a,a',b,',c,c'=-2}^2 a_{\kappa,a}a_{\kappa,a'} e^{-E_\kappa t} \sum_{\alpha, \beta} \sum_{m,n} \frac{\delta_{b-c, b'-c'}}{1 - \mu_c^\alpha} \sqrt{p_\alpha^{eq} p_\beta^{eq}}\tilde{X}_\alpha^{(n,b-c,m)} \tilde{X}_\beta^{(n,b-c,m)} e^{\lambda_{n,b-c,m} t} \times &\\
\frac{I_{b,c}^n(\theta_c^\alpha) I_{b',c'}^n(\theta_c^\beta)}{\sqrt{H_{b-c,n}(\mu_c^\alpha) H_{b-c,n}(\mu_c^\beta)}} \mathcal{D}_{a,b}^*(\Omega_{J,W_\alpha}) \mathcal{D}_{a',b'}(\Omega_{J,W_\beta}) \mathcal{D}_{c,0}^*(\Omega_{SF,i}) \mathcal{D}_{c',0}(\Omega_{SF,j}). & 
	\end{aligned}
\end{equation}
Such a model can be computationally demanding and might be difficult to use on experimental data where the contribution from wobbling motions is expected to be small.

%% file: Section/5_CSAjumpRelaxation.tex
\section{Time-dependent interaction strengths}
When we introduced the BWR relaxation theory, we implicitly assumed that the strengths of the interactions were time independent. Chemical bonds vibrate in the femtosecond range by a few picometers, so that it is a valid approximation in the case of the DD interactions between directly bonded nuclei, even if its effective strength needs to be carefully set \cite{Case_JBNMR_1999}. Similarly, bonds vibrations lead to slight variations in CSA tensors for the nitrogen and carbonyl carbon-13 of peptide planes \cite{Tang_JBNMR_2007}. S.\,Tang and D.\,Case indicate that the use of a scaling factor for the CSA when analyzing relaxation data is a way to take into account the motional averaging of the CSA tensor \cite{Tang_JBNMR_2007}. However, to the best of our knowledge, no study focused on the variations of CSA tensors in protein side-chains and their effect on relaxation. CSA tensors have been measured by solid-state NMR \cite{liang_chemrev_2022} and shown to be conformation dependent, particularly for alpha and beta carbon-13 nuclei \cite{hong_jacs_2000,wylie_jacs_2005}. Rienstra and coworkers showed that CSA tensors could be used as efficient constraints to determine the structure of a protein \cite{wylie_jacs_2009,wylie_pnas_2011}. \\
We restrict ourselves here to the simple case of a methyl group exchanging between two rotamers in a protein undergoing isotropic overall diffusion with correlation time $\tau_c = 10$\,ns. We assume perfect tetrahedral geometry for the methyl group, and a diffusion coefficient for the rotation $D_{rot} = 5 \times 10^{10}$\,s$^{-1}$. The set of Euler angles for transformation from the jump frame to the rotamer frame is $\Omega_{J,R} = \{\pm \pi/2, \beta_J, 0\}$ with $\beta_J = 76$\,deg, which is typical for carbon chains. The carbon-13 CSA tensors are considered axially symmetric and aligned along the C-C bond of the methyl group. The CSA of proton is neglected  \cite{Tugarinov_JBNMR_2004}. We set the exchange rate from state 1 to 2 to $k_{21} = 0.5 \times 10^9$\,s$^{-1}$ and the exchange rate $k_{21}$ is calculated to satisfy the microreversibility condition (Eq.\,\ref{eq:MicroRev}) depending on the three situations we consider for the population of state 1: $p_1=0.3$, $p_1=0.5$ and $p_1=0.7$. Finally, the population average of the axially symmetric carbon-13 CSA is set to $\sigma_{av}=18.2$\,ppm which corresponds to the value of the carbon-$\delta1$ in isoleucine side-chains determined from cross-correlated cross-relaxation rates \cite{Tugarinov_JBNMR_2004}. We evaluate the effect of $\Delta \sigma = \sigma_2 - \sigma_1$, the difference in CSA between the two states by focusing on carbon-13 longitudinal and transverse relaxation rates and the $^{13}$C-CSA/DD cross-correlated cross-relaxation rates $\eta_z$ and $\eta_{xy}$ at a magnetic field $B_0=14.1$\,T \cite{BolikCoulon_JMR_2020}:
\begin{equation}
	\begin{aligned}
R_1 (^{13}\mathrm{C}) =& \frac{1}{4} (\mathcal{J}_{CH}^{(I)} (\omega_H - \omega_C) + 3 \mathcal{J}^{(I)}_{CH} (\omega_C) + 6 \mathcal{J}^{(I)}_{CH} (\omega_H + \omega_C)) \\
		& + \frac{4}{3} (\mathcal{J}^{(I)}_{CD} (\omega_C - \omega_D) + 3 \mathcal{J}^{(I)}_{CD}(\omega_C) + 6\mathcal{J}^{(I)}_{CD}(\omega_C+\omega_D)) + \mathcal{J}^{(I)}_{C} (\omega_C), \\
R_2 (^{13}\mathrm{C}) =& \frac{1}{8} (4 \mathcal{J}^{(I)}_{CH}(0) + \mathcal{J}^{(I)}_{CH} (\omega_H - \omega_C) + 3 \mathcal{J}^{(I)}_{CH}(\omega_C) + 6 \mathcal{J}^{(I)}_{CH}(\omega_H) +  6 \mathcal{J}^{(I)}_{CH}(\omega_H + \omega_C)) \\
				&+ \frac{2}{3}(4 \mathcal{J}^{(I)}_{CD}(0) + \mathcal{J}^{(I)}_{CD} (\omega_C - \omega_D) + 3 \mathcal{J}^{(I)}_{CD}(\omega_C) + 6 \mathcal{J}^{(I)}_{CD}(\omega_D) + 6 \mathcal{J}^{(I)}_{CD}(\omega_C + \omega_D)) \\
				&+ \frac{1}{6} (4 \mathcal{J}^{(I)}_C (0) + 3 \mathcal{J}^{(I)}_C (\omega_C)), \\
\eta_z (^{13}\mathrm{C}) =& \sqrt{\frac{3}{2}} \mathcal{J}^{(I)}_{C,CH}(\omega_C), \\
\eta_{xy} (^{13}\mathrm{C}) =& \frac{1}{2} \sqrt{\frac{1}{6}} (4 \mathcal{J}^{(I)}_{C,CH}(0) + 3 \mathcal{J}^{(I)}_{C,CH}(\omega_C)),
	\end{aligned}
\end{equation}
where the superscript $(I)$ indicates that the strengths of the interactions are included in the spectral density functions which are written:
\begin{equation}
	\begin{aligned}
&\mathcal{J}^{(I)}_{CH} (\omega) = d_{CH}^2 \frac{2}{5}  \sum_{a,b=-2}^2  \sum_{\alpha,\beta=1}^2 \sum_{n=0}^1 \frac{\tau_{b,n}}{1+(\omega \tau_{b,n})^2} \sqrt{p_\alpha p_\beta} \tilde{X}_\alpha^{(n)} \tilde{X}_\beta^{(n)}  \mathcal{D}_{a,b}^{(2)*}(\Omega_{J,R_\alpha}) \mathcal{D}_{a,b}^{(2)}(\Omega_{J,R_\beta}) d_{b,0}(\beta_{CCH})^2, \\
&\mathcal{J}^{(I)}_{CD} (\omega) = d_{CD}^2 \frac{2}{5}  \sum_{a,b=-2}^2  \sum_{\alpha,\beta=1}^2 \sum_{n=0}^1 \frac{\tau_{b,n}}{1+(\omega \tau_{b,n})^2} \sqrt{p_\alpha p_\beta} \tilde{X}_\alpha^{(n)} \tilde{X}_\beta^{(n)}  \mathcal{D}_{a,b}^{(2)*}(\Omega_{J,R_\alpha}) \mathcal{D}_{a,b}^{(2)}(\Omega_{J,R_\beta}) d_{b,0}(\beta_{CCH})^2, \\
&\mathcal{J}^{(I)}_{C} (\omega) = \frac{2}{3} \omega_C^2  \frac{2}{5} \sum_{\alpha,\beta=1}^2 \sum_{n=0}^1 \sigma_\alpha \sigma_\beta \frac{\tau_{0,n}}{1+(\omega \tau_{0,n})^2} \sqrt{p_\alpha p_\beta} \tilde{X}_\alpha^{(n)} \tilde{X}_\beta^{(n)} \mathcal{P}_2(\cos \theta_{\alpha,\beta}) , \\
&\mathcal{J}^{(I)}_{C,CH} (\omega) = d_{CH} \sqrt{\frac{2}{3}} \omega_C \frac{2}{5} \mathcal{P}_2(\cos \beta_{CCH})  \sum_{\alpha, \beta=1}^2 \sum_{n=0}^1 \sigma_\alpha \frac{\tau_{0,n}}{1+(\omega \tau_{0,n})^2} \sqrt{p_\alpha p_\beta} \tilde{X}_\alpha^{(n)} \tilde{X}_\beta^{(n)}  \mathcal{P}_2(\cos \theta_{\alpha,\beta}),
	\end{aligned}
\end{equation}
where $d_{CX}  = -\mu_0 \hbar \gamma_C \gamma_X / (4\pi r_{CX}), X={H, D}$  with $\mu_0$ the permeability of free space, $\hbar$  the Planck's constant devided by $2\pi$, $\gamma_A$ the gyromagnetic ratio of nucleus $A$ and $r_{CX}$ the distance between the $^{13}$C and nucleus X,  $\beta_{CCH}=180-109.47=70.53$\,deg is the angle between the C-H bond and the methyl group symmetry axis, $\theta_{\alpha,\beta}$ is the angle between the vectors pointing along the directions of the C-C bonds in rotamers $\alpha$ and $\beta$ (that is $\theta_{\alpha,\beta} =0$ when $\alpha=\beta$ and $\theta_{\alpha \beta} = 2\beta_{CC}$ when $\alpha \neq \beta$), and:
\begin{equation}
\tau_{b,n}^{-1} = \tau_c^{-1} - \lambda_n + b^2 D_{rot},
\end{equation}
where $\lambda_n$ is the $n^{th}$ eigenvalue of the symmetrized exchange matrix.
\begin{figure*}[!ht]
	\begin{center}
		\includegraphics[width=0.9\textwidth]{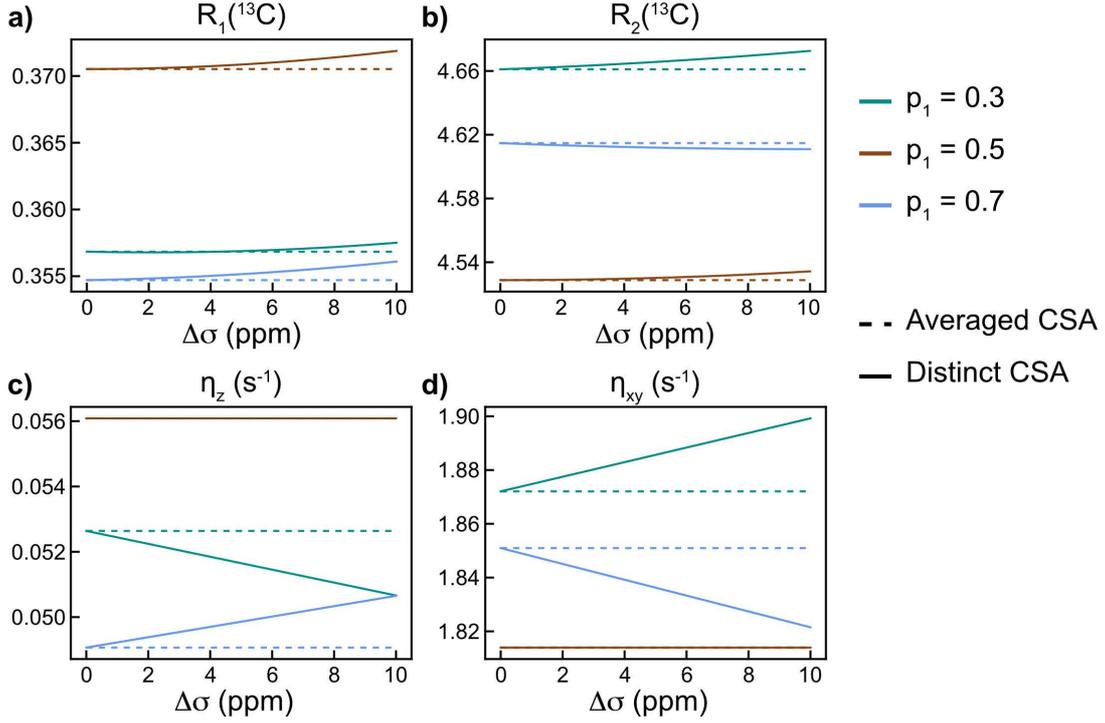}
	\end{center}
	\caption{Evolution of the carbon-13 longitudinal R$_1$ (\textbf{a}) and transverse R$_2$ (\textbf{b}) relaxation rates, as well as carbon longitudinal (\textbf{c}) and transverse (\textbf{d}) cross-correlated cross-relaxation rate for a methyl group exchanging between two rotamer positions as a function of the difference between the CSA of the two rotamers. Calculations are performed for three equilibrium position for the state 1 and by either considering a population-averaged CSA value (dash) or distinct CSA tensors (solid).}
	\label{fig:CSArotRelax2States}
\end{figure*}
The carbon auto-relaxation rates show very small deviations depending on whether distinct CSA values are considered or not (Fig.\,\ref{fig:CSArotRelax2States}a,b), which most likely arises from the relatively small contribution of the CSA (about 1.5\,\% for the carbon-R$_1$ and 5\,\% for the carbon-R$_2$) compared to the DD interactions. The cross-correlated cross-relaxation rates calculated with distinct CSA tensors show a linear variation with the difference in anisotropy between the two states (Fig.\,\ref{fig:CSArotRelax2States}c,d). This can be understood by expanding the spectral density function for cross-correlation between the carbon-CSA and carbon-proton DD interactions:
\begin{equation}
	\begin{aligned}
\mathcal{J}^{(I)}_{C,CH}(\omega) =& d_{CH} \sqrt{\frac{2}{3}} \omega_C \frac{2}{5} \mathcal{P}_2(\cos \beta_{CCH}) \left[\sigma_{av} \left( \frac{\tau_c}{1 + (\omega \tau_c)^2} (p_1^2 + p_2^2 + 2 p_1 p_2 \mathcal{P}_2 (\cos \beta_J))    \right. \right. \\
	& \left. \left. + \frac{2 \tau_{0,1}}{1 + (\omega \tau_{0,1})^2} p_1 p_2 (1 - \mathcal{P}_2 (\cos \beta_J)) \right) \right. \\
	& \left. + \Delta \sigma p_1 p_2 (p_1 - p_2) (1 - \mathcal{P}_2 (\cos \beta_J)) \left(\frac{\tau_{0,1}}{1 + (\omega \tau_{0,1})^2} - \frac{\tau_c}{1 + (\omega \tau_c)^2} \right) \right],
	\end{aligned}
\label{eq:JCrossCorr2States}
\end{equation}
where the last line contains the $\Delta \sigma$-dependent part of the spectral density function and highlights the linear variation of the cross-correlated cross-relaxation rates when the difference of CSA is changed. For the chosen geometry, we have:
\begin{equation}
	\begin{aligned}
& d_{CH} \mathcal{P}_2 (\cos \beta_{CCH}) > 0, \\
& 1 - \mathcal{P}_2 (\cos \beta_J) > 0.
	\end{aligned}
\end{equation}
In addition, we have chosen $\Delta \sigma  = \sigma_2 - \sigma_1 > 0$ and:
\begin{equation}
	\begin{aligned}
& \tau_{0,1} - \tau_c = - \tau_c^2 \frac{k_{12} + k_{21}}{1 + \tau_c (k_{12} + k_{21})} < 0, \\
& \frac{\tau_{0,1}}{1 + (\omega_C \tau_{0,1})^2} - \frac{\tau_c}{1 + (\omega_C \tau_c)^2} > 0,
	\end{aligned}
\end{equation}
such that, when calculations are performed with distinct CSA tensors, $\eta_z$ increases when $p_1 > p_2$ and decreases otherwise, and since $J^{(I)}_{C,CH}(0) \gg J^{(I)}_{C,CH} (\omega_C)$, $\eta_{xy}$ decreases when $p_1 > p_2$ and increases otherwise. Finally, when $p_1 = p_2$, the $\Delta \sigma$-dependent term in Eq.\,\ref{eq:JCrossCorr2States} vanishes and the cross-correlated cross-relaxation rates are independent from the difference in CSA value between the two states. \\
Here, we used a simple model to highlight the contribution of internal dynamics to relaxation when a spin system exchanges between discrete positions with distinct CSA tensors. Calculations show that small effects on relaxation rates can be expected from changes of the amplitude of the CSA of less than 10 ppm between conformers (Fig.\,\ref{fig:CSArotRelax2States}).  Investigations combining solid-state NMR, relaxation and density functional theory calculations of CSA parameters may offer experimental evidence of the contribution of such mechanisms to relaxation.

%% file: Section/6_Discussion.tex
\section{Discussion}
The models presented in the previous sections all share one property: they are complex. Up to 3 correlation times might be necessary to model the global tumbling; the number of decaying exponentials describing rotamer jumps increases linearly with the number of accessible states; modeling the wobbling in a cone motion involves evaluating difficult integrals. Out of the 4 considered motions, the rotation on a cone seems the most simple one. In addition to their intrinsic complexity, another difficulty arises when analyzing data, in particular NMR relaxation rates: how to choose one model over the other without \textit{a priori} knowledge of the type of motions involved? These aspects naturally led to the popularization of the Model-Free correlation function \cite{Lipari_JACS_1982_1} to model internal motions:
\begin{equation}
C_{MF}^{(i)}(t) = \mathcal{S}_{MF}^2 + (1 - \mathcal{S}_{MF}^2)e^{-t/\tau_{MF}},
\label{eq:MFcorrFunc}
\end{equation}
where $\mathcal{S}_{MF}^2$ is the squared order parameter and $\tau_{MF}$ an effective correlation time. The simplicity of this model makes it particularly attractive, and has indeed led to a vast number of successful analyses of 
NMR relaxation experiments \cite{Lipari_JACS_1982_2, Schneider_Biochemistry_1992, Bruchweiler_Science_1995, Tjandra_JACS_1995, Mandel_JMB_1995, Farrow_Biochemistry_1995, Lee_JBNMR_1997, Hall_JACS_2006, Miloushev_Structure_2008, Liao_JPCB_2012}. Over the past few decades, the MF approach has been modified in order to include the effect of cross correlation \cite{Kay_JMR_1991}, or to include more than one correlation time and study more complex systems \cite{Clore_JACS_1990,Calandrini_JCP_2008,Calandrini_JCP_2010,Calligari_JPCB_2012}, in particular intrinsically disordered proteins \cite{Buevich_JACS_1999,Buevich_JBNMR_2001,Khan_BiophysJ_2015,Hsu_Biophys_2018,Adamski_JACS_2019}. However, a number of questions arises on this simple form of correlation function, and in particular regarding the factorization of the correlation functions accounting for internal motions (Eq.\,\ref{eq:MFcorrFunc}) and global tumbling. It is clear from section\,\ref{section:GlobalTumblingCorrFunc} that such factorization is mathematically exact only in the case of an isotropic diffusion tensor, which is the frame in which the MF approach was initially presented. An approximated form for the global tumbling correlation function in the case of an axially symmetric diffusion tensor was proposed as well, and consists in the sum of two decaying exponential, with decay constants and relative weights that can be determined experimentally \cite{Lipari_JACS_1982_1,Dosset_JBNMR_2000,Walker_JMR_2004,	Auvergne_JBNMR_2008}. Then, how is the factorization of the global tumbling and internal motion correlation functions affecting the fitted values of $\mathcal{S}_{MF}^2$ and $\tau_{MF}$? It must be noted that the SRLS model of correlation function was introduced in part as an attempt to overcome this difficulty \cite{Polimeno_JPC_1995,Tugarinov_JACS_2001,Meirovitch_JPCA_2006}. Here, we evaluate the effect of the factorization of the global tumbling correlation function on the value of the fitted MF parameters. We focus on auto-correlation only:
\begin{equation}
 C(t) = \sum_{a,a',b,b'=-2}^2 \langle \mathcal{D}_{q,a}^{(2)*}(\Omega_{L,D},0) \mathcal{D}_{q,a'}^{(2)}(\Omega_{L,D},t) \rangle \mathcal{D}_{a,b}^{(2)*} (\Omega_{D,M}) \mathcal{D}_{a'b'}^{(2)}(\Omega_{D,M} \langle \mathcal{D}_{b,0}^{(2)*}(\Omega_{M,i},0) \mathcal{D}_{b',0}^{(2)}(\Omega_{M,i}, t) \rangle, 
\end{equation}
where $M$ denotes the frame in which the motion is best described, $\langle \mathcal{D}_{q,a}^{(2)*}(\Omega_{L,D},0) \mathcal{D}_{q,a'}^{(2)}(\Omega_{L,D},t) \rangle$ accounts for overall rotational diffusion, $\Omega_{D,SF}$ is the angle orienting the diffusing frame in the global diffusion tensor frame, and $\langle \mathcal{D}_{b,0}^{(2)*}(\Omega_{M,i},0) \mathcal{D}_{b',0}^{(2)}(\Omega_{M,i}, t) \rangle$ accounts for internal motions. The MF correlation function is written as:
\begin{equation}
C_{MF}(t) =   \sum_{a,a'=-2}^2  \langle \mathcal{D}_{q,a}^{(2)*}(\Omega_{L,D},0) \mathcal{D}_{q,a'}^{(2)}(\Omega_{L,D},t) \rangle  \mathcal{D}_{a,0}^{(2)*} (\Omega_{D,M}) \mathcal{D}_{a'0}^{(2)}(\Omega_{D,M})  \left(\mathcal{S}^2_{MF} + (1 - \mathcal{S}^2_{MF} ) e^{-t/\tau_{MF}} \right) .
\label{eq:GenMFCorrFunc}
\end{equation}
In the case of an isotropic diffusion tensor, the factorization of the global tumbling correlation function is exact and the correlation functions are written:
\begin{equation}
	\begin{aligned}
&C(t) = \frac{1}{5} e^{-t/\tau_c} \sum_{b=-2}^2 \langle \mathcal{D}_{b,0}^{(2)*}(\Omega_{M,i},0) \mathcal{D}_{b,0}^{(2)}(\Omega_{M,i}, t) \rangle, \\
&C_{MF}(t) = \frac{1}{5} e^{-t/\tau_c} \left(\mathcal{S}^2_{MF} + (1 - \mathcal{S}^2_{MF} ) e^{-t/\tau_{MF}} \right).
	\end{aligned}
\end{equation}
\begin{table*}[t]
	\caption{Values of the fitted MF parameters (correlation function Eq.\,\ref{eq:GenMFCorrFunc}) for 4 types of internal motions and the three possible symmetry properties of the overal diffusion tensor. In the case of the symmetric top and fully asymmetric overal diffusion tensor, we report the average and standard deviation over the possible orientations, \textit{i.e.} the different values of $\theta_{D,M}$ and $\{ \varphi_{D,M},\theta_{D,M} \}$ respectively. In these fits, the values and orientation of the diffusion tensor (angle $\Omega_{D,M}$) were fixed such that $\mathcal{S}_{MF}^2$ and $\tau_{MF}$ were the only adjustable parameters. The diffusion tensor for global tumbling has $D_{xx}=1.64 \times 10^7$\,s$^{-1}$, $D_{yy}=1.82\times10^7$\,s$^{-1}$ and $D_{zz}=2.35 \times 10^7$\,s$^{-1}$ which corresponds to the diffusion tensor for the proximal Ubiquitin in a diubiquitin with linkage at lysine-11 \cite{Castaneda_PCCP_2016}. When the symmetric model is used, eigenvalues are $D_\parallel = D_{zz}$ and $D_\perp = \frac{1}{2} (D_{xx} + D_{yy})$. For the isotropic diffusion, the eigenvalue equals $D=\frac{1}{3}(D_{xx}+D_{yy}+D_{zz})$. The orientation of the rotamer frames in the jump frame are $\Omega_{J,R}^{(2)}=\{\pm\frac{\pi}{2},\beta_{CC}\}$ when two states are considered, and $\Omega_{J,R}^{(3)}=\{2n\frac{\pi}{3},\beta_{CC}\}$ for the 3-state jump with $n=0, 1, 2$ and where $\beta_{CC}=76^\circ$. The 3-state jump model with unequal populations (bottom row) was simulated with $p_1 = 0.7$, $p_2 = 0.2$ and $p_3=0.1$. The rotation on a cone is modeled assuming a methyl group gegometry and with $D_{rot}=10^{11}$\,s$^{-1}$. For the wobbling motion, the cone semi-angle opening is $\beta_c = 20^\circ$ and $D_W = 10^8$\,s$^{-1}$.}
	\begin{center}
		{\def\arraystretch{1.5}
		\begin{tabular}{|c|c|c|c|c|c|c|}
			\hline
		 	 & \multicolumn{2}{c|}{Isotropic} & \multicolumn{2}{c|}{Symmetric top} & \multicolumn{2}{c|}{Fully asymmetric} \\%
			\hline
			& $\mathcal{S}_{MF}^2$ & $\tau_{MF}$ (ps) & $\mathcal{S}_{MF}^2$  & $\tau_{MF}$ (ps)& $\mathcal{S}_{MF}^2$ & $\tau_{MF}$ (ps)\\ 
			\hline
			rotation & $0.12$ & $4.8$ & $0.12 \pm 0.00$ & $4.8 \pm 0.0$ & $0.12 \pm 0.00$ & $4.8 \pm 0.0$ \\
			wobbling & $0.83$ & $339$ &$0.83 \pm 0.00$ & $344 \pm 5$ & $0.83 \pm 0.00$ & $344 \pm 5$ \\
			2-state jump &  \multirow{2}{*}{$0.83$} &  \multirow{2}{*}{$471$} &  \multirow{2}{*}{$0.77 \pm 0.06$} &  \multirow{2}{*}{$1,290 \pm 709$} &  \multirow{2}{*}{$0.77 \pm 0.06$} &  \multirow{2}{*}{$1,319 \pm 748$} \\
			$p_1=p_2$ & & & & & & \\
			3-state jump & \multirow{2}{*}{$0.17$} & \multirow{2}{*}{$336$} & \multirow{2}{*}{$0.17 \pm 0.00$} & \multirow{2}{*}{$333 \pm 3.0$} & \multirow{2}{*}{$0.17 \pm 0.00$} & \multirow{2}{*}{$333 \pm 3$} \\
			$p_1=p_2=p_3$ & & & & & & \\
			\hline
			\hline
			3-state jump & \multirow{2}{*}{$0.43$} & \multirow{2}{*}{$157$} & \multirow{2}{*}{$0.43 \pm 0.03$} & \multirow{2}{*}{$161 \pm 55$} & \multirow{2}{*}{$0.43 \pm 0.03$} & \multirow{2}{*}{$164 \pm 57$} \\
			$p_1 \neq p_2 \neq p_3$ & & & & & &  \\
			\hline
		\end{tabular}
		}
	\end{center}
	\label{table:FitParamMF}
\end{table*}
We considered the diffusion tensor of diubiquitin with linkage at lysine-11 \cite{Castaneda_PCCP_2016}, which has anisotropy of 1.36 and rhombicity 0.44. The fitted parameters for internal dynamics (Table\,\ref{table:FitParamMF}) can be used as reference for the parameters fitted in the case of an asymmetric overall diffusion tensor: if the factorization  does not affect the value of the fitted parameters, we would expect to find the same values for different tensor symmetries. \\
For asymmetric diffusion tensor, the factorization of the global tumbling correlation function is not mathematically exact. The two correlation functions for the symmetrical top diffusion tensor read \cite{Lipari_JACS_1982_1, Barbato_Biochemistry_1992}:
\begin{equation}
	\begin{aligned}
&C(t) = \frac{1}{5} \sum_{a,b,b'=-2}^2 e^{-(6 D_\perp + a^2 (D_\parallel - D_\perp))t} d_{a,b}(\theta_{D,M})   d_{a,b'}(\theta_{D,M}) \langle \mathcal{D}_{b,0}^{(2)*}(\Omega_{M,i},0) \mathcal{D}_{b,0}^{(2)}(\Omega_{M,i}, t) \rangle, \\
&C_{MF}(t) =  \frac{1}{5} \left(\mathcal{S}^2_{MF} + (1 - \mathcal{S}^2_{MF} ) e^{-t/\tau_{MF}} \right) \sum_{a=-2}^2 e^{-(6 D_\perp + a^2 (D_\parallel - D_\perp))t} \left[d_{a,0}(\theta_{D,M}) \right]^2,
	\end{aligned}
\end{equation}
and for the fully asymmetric tensor \cite{Tjandra_JACS_1995}:
\begin{equation}
	\begin{aligned}
& C(t) = \frac{1}{5} \sum_{a,a',b,b'=-2}^2 \sum_{\kappa=1}^5 a_{\kappa,a} a_{\kappa,a'} e^{-E_\kappa t}    e^{i \varphi_{D,M} (a - a')} d_{a,b}(\theta_{D,M}) d_{a,b'}(\theta_{D,M})\langle \mathcal{D}_{b,0}^{(2)*}(\Omega_{M,i},0) \mathcal{D}_{b,0}^{(2)}(\Omega_{M,i}, t) \rangle, \\
& C_{MF}(t) =  \frac{1}{5} \left(\mathcal{S}^2_{MF} + (1 - \mathcal{S}^2_{MF} ) e^{-t/\tau_{MF}} \right) \sum_{a,a'=-2}^2 \sum_{\kappa=1}^5 a_{\kappa,a} a_{\kappa,a'} e^{-E_\kappa t}  e^{i \varphi_{D,M} (a - a')} \left[d_{a,0}(\theta_{D,M})\right]^2.
	\end{aligned}
\end{equation}
We can note here that the MF correlation function cannot distinguish the orientation of the interaction frames in the diffusion frame, which is particularly critical in the presence of a rotamer jump motion. The fitted MF parameters agree well with the ones obtained when simulating an isotropic global tumbling, except in the case of the two-state jump model where large deviations of the fitted parameters are obtained when $\Omega_{D,M}$ is changed (Table\,\ref{table:FitParamMF}). Two- and three-state jump models have been studied extensively by solid-state deuterium NMR, based on the investigation of linewidths and relaxation rates \cite{gall_jacs_1981,rice_jacs_1981,batchelder_jacs_1983,Straus_JBNMR_1997,vugmeyster_methods_2018}. J.\,Wand and co-workers already distinguished the 2-state and 3-state jump models by considering the resulting symmetry \cite{Frederick_JCPB_2008}. In our case of 3 exchanging states with equal populations, the motion is azymutal symmetric, that is:
\begin{equation}
\langle x^2 \rangle = \langle y^2 \rangle,\,\,\,\langle x \rangle = \langle y \rangle = 0,\,\,\,\mathrm{and}\,\,\, \langle x y \rangle = 0,
\label{eq:AymSymConditions}
\end{equation}
where $\langle \cdot \cdot \cdot \rangle$ denotes an ensemble average, and $x$ and $y$ are the $x$- and $y-$coordinates of the interaction frame main axis in the jump frame (labelled $M$ in this section). For the 2-state jump model presented here, the first condition in Eq.\,\ref{eq:AymSymConditions} is not fulfilled. It is not fulfilled neither when populations in the 3-state jump model are not equal, in which case the factorization of the asymmetric global tumbling correlation functions does not accurately reproduces MF parameters obtained in the isotropic tumbling case, and show large deviations depending on the orientation in the diffusion tensor frame (Table\,\ref{table:FitParamMF}). These conclusions also apply for higher asymmetry of the diffusion tensor (Table\,S1). \\
These simulations suggest that the factorization of the global tumbling correlation function does not alter the value of the fitted parameters in the presence of a diffusing-type of internal motions (that is diffusion on a cone or wobbling in a cone). When the sampling of the conformational space is discrete (jump model), the factorization does not affect the value of the MF fitted parameters only when the motion has azymutal symmetry, a strong condition that may not be met in most cases when studying protein side-chains dynamics. Overall, explicit models of motions are particularly powerful to account for the orientation of interaction frames in the molecular frame. This can prove itself especially adapted in the case of non-axially symmetric CSA tensors. Their contribution to relaxation can be decomposed in contribution from two orthogonal axially symmetric tensors which orientation in the molecular frame are different and can easily be included in the explicit models of motions.

%% file: Section/7_Conclusion.tex
\section{Conclusion}
We have reviewed explicit models of motions that can be relevant to study the dynamic properties of biomolecules. For the past 40 years, the Model Free approach has been prefered for the analysis of NMR relaxation data as it does not require \textit{a priori} knowledge on the nature of motions of the bond vectors. The advances in MD simulations methods can now provide such information, for instance with the help of the detector analysis which can easily distinguish the contribution of motions originating from different times-scales \cite{Smith_JCP_2018,Smith_JCP_2019,Smith_anie_2019,Smith_JBNMR_2021,Zumpfe_FMB_2021}. Explicit models of motions as presented here can be used to obtain a mechanistic picture of the motions from a combined NMR and MD analysis \cite{Smith_JMRO_2022,BolikCoulon_ArXiv_2022}. We finally discussed the use of MF when analyzing NMR relaxation data when the overall diffusion tensor is anisotropic. In this case, the MF fitted parameter are affected by the orientation of the interaction tensor frames in the overall diffusion tensor frame, in particular when the motion is not azymutal symmetric. We expect this theoretical work, in pair with softwares for the calculation of the relaxation rates of nuclear spins \cite{kuprov_jmr_2007,bengs_mrc_2018,BolikCoulon_JMR_2020}, to support the development of system-specific explicit models of motions for the analysis of NMR relaxation.

%% file: Section/SI.tex
\newpage
\section*{Supplementary Material}
\beginsupplement

\begin{figure}[!ht]
	\begin{center}
		\includegraphics[width=0.9\textwidth]{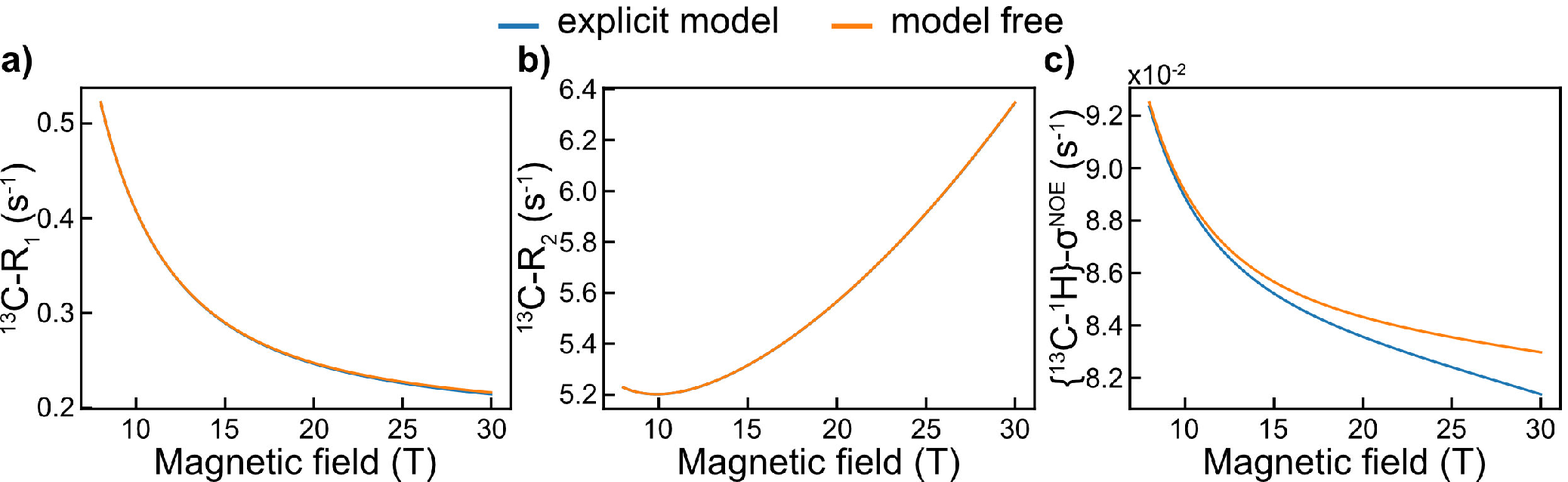}
	\end{center}
	\caption{Carbon longitudinal R$_1$ (\textbf{a}) and transverse R$_2$ relaxation rates (\textbf{b},), and carbon-proton cross-relaxation rate $\sigma^\mathrm{NOE}$ (\textbf{c}) at 14.1\,T in a $^{13}$C$^1$H$^2$H$_2$ methy group for a correlation time for isotropic global tumbling $\tau_c=10$\,ns shown as a function of the magnetic field. Two models of motions were used, each of them only considering the methyl rotation: rotation on a cone (blue) and model free (orange). The angle between the C-C and C-H bonds is fixed to $109.47^\circ$. The diffusion coefficient for methyl rotation is set to $D_{rot}=5\times10^{10}$\,s$^{-1}$ leading to a correlation time of 10\,ps.}
	\label{fig:ComparisonRotMF}
\end{figure}

\begin{figure}[!ht]
	\begin{center}
		\includegraphics[width=0.9\textwidth]{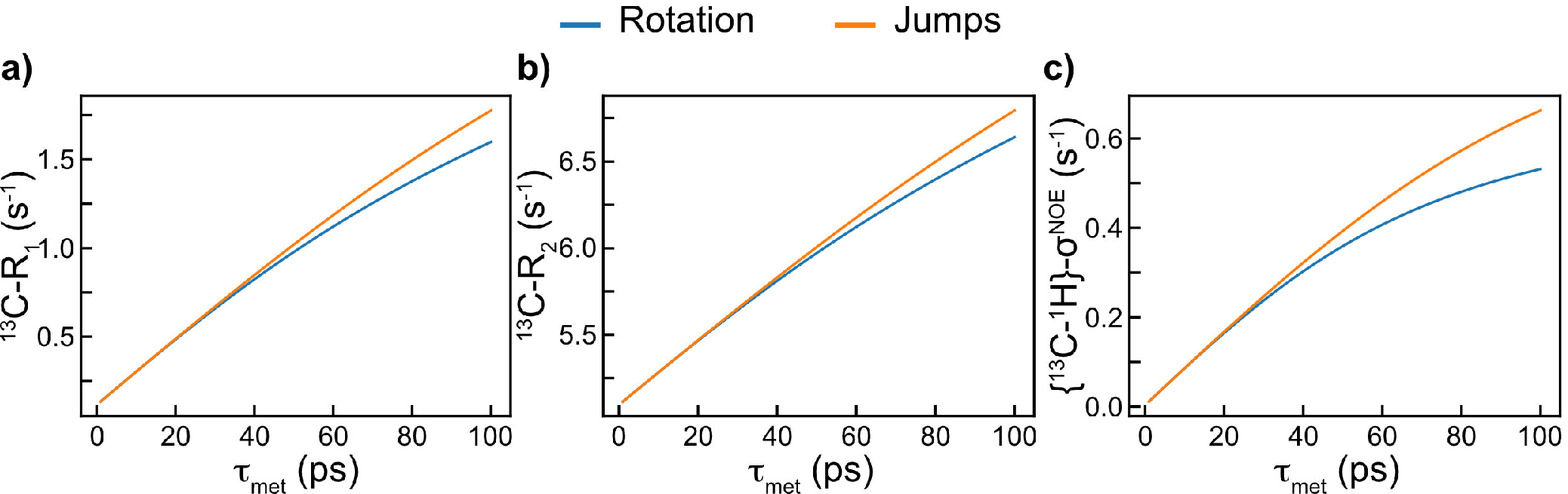}
	\end{center}
	\caption{Carbon longitudinal R$_1$ (\textbf{a}) and transverse R$_2$ relaxation rates (\textbf{b},), and carbon-proton cross-relaxation rate $\sigma^\mathrm{NOE}$ (\textbf{c}) at 14.1\,T in a $^{13}$C$^1$H$^2$H$_2$ methy group for a correlation time for isotropic global tumbling $\tau_c=10$\,ns shown as a function of the correlation time for methyl rotation. Two models of motions were used, each of them only considering the methyl rotation: rotation on a cone (blue) and jumps (orange). The angle between the C-C and C-H bonds is fixed to $109.47^\circ$. In the case of jumps between discrete position, the angle between each of them is fixed to $2\pi/3$ and the populations are supposed to be equal for the three conformers. The correlation time for methyl rotation is calculated as $\tau_{met} = \frac{1}{1 - \mathcal{S}^2_m} \sum_{a=-2}^2 \sum_{\alpha, \beta = 1}^3 \sum_{n,\lambda_n != 0} -\frac{\sqrt{p_\alpha p_\beta}}{\lambda_n} \tilde{X}_\alpha^{(n)} \tilde{X}_\beta^{(n)} \mathcal{D}_{a,0}^{(2)*} (\Omega_{CCH}) \mathcal{D}_{a,0}^{(2)}(\Omega_{CCH})$ where $\mathcal{S}^2_m$ is the order parameter for methyl rotation (identical for both models), $p_\alpha=1/3$ is the fractional population for conformer $\alpha$, $\tilde{X}_\alpha^{(n)}$ is the value $\alpha$ for the eigenvector of the exchange matrix associated to the eigenvalue $\lambda_n$ and $\Omega_{CCH}$ is the Euler angle for transformation from the C-C to C-H bond. The jump rate between each conformer is equal to $k_{ex}$ and the associated diffusion constant for methyl rotation used in the rotation on a cone model is calculated using Eq.\,76 of the main text.}
	\label{fig:ComparisonRotEx}
\end{figure}

\begin{figure}[!ht]
	\begin{center}
		\includegraphics[width=0.9\textwidth]{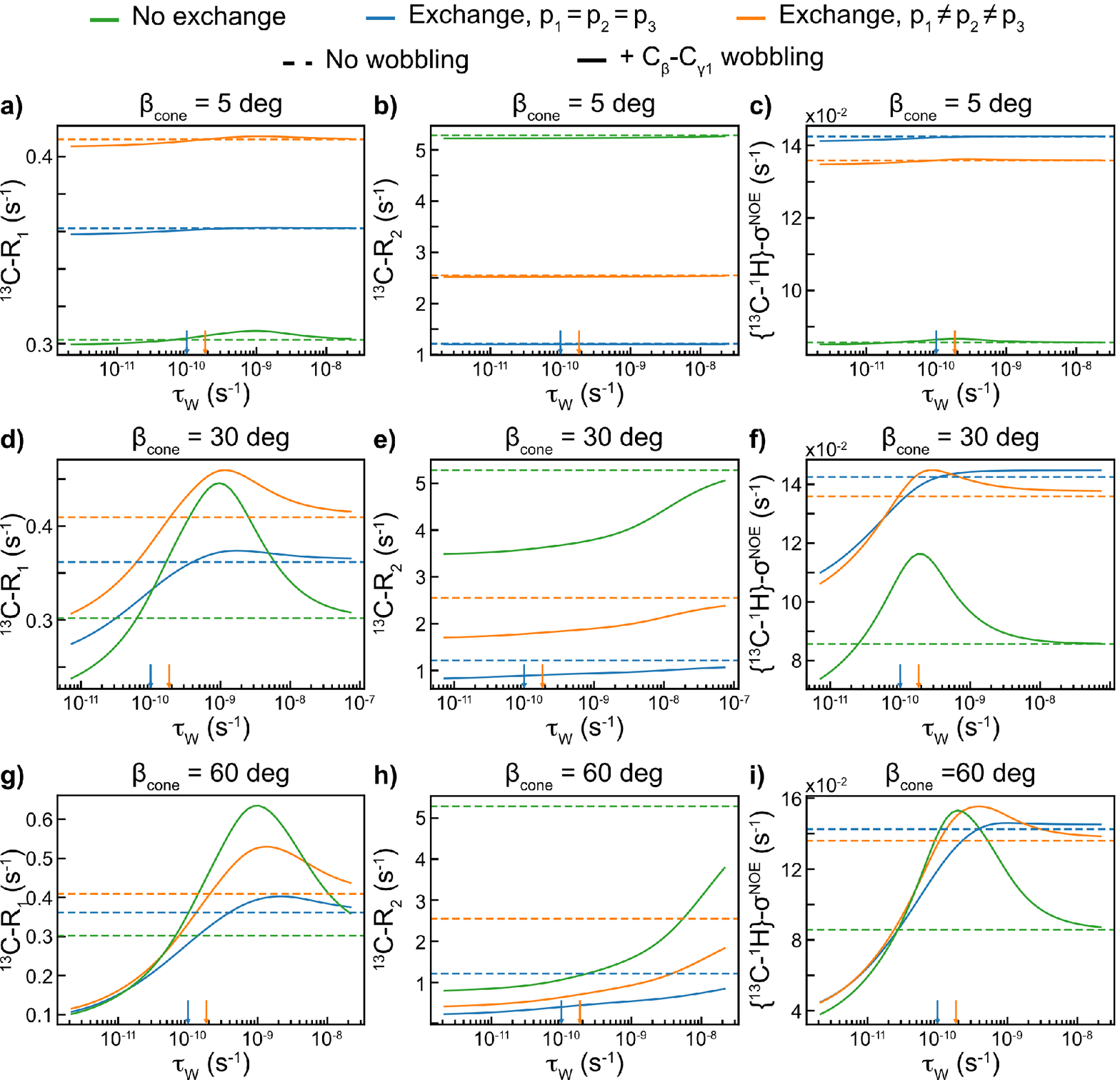}
	\end{center}
	\caption{Contribution of rotamer jump and wobbling in a cone motions to relaxation rates. Carbon longitudinal R$_1$ (\textbf{a}, \textbf{d}, \textbf{g}) and transverse R$_2$ relaxation rates (\textbf{b}, \textbf{e}, \textbf{h}), and carbon-proton cross-relaxation rate $\sigma^\mathrm{NOE}$ (\textbf{c}, \textbf{f}, \textbf{i}) at 14.1\,T in a $^{13}$C$^1$H$^2$H$_2$ methy group for a correlation time for isotropic global tumbling $\tau_c=10$\,ns. Dashed horizontal lines show the value of the relaxation rates without wobbling. Relaxation rates calculated in the abscence of rotamer jumps are shown in green. In the presence of rotamer jumps, the case where all rotamer populations are equal (blue) and unequal (orange) are distinguished. In this latter case, populations are $p_1=0.7$, $p_2=0.2$ and $p_1=0.1$. Calculations are shown as a function of the correlation time for wobbling, a function of the wobbling diffusion constant and cone semi-angle opening $\beta_{c}$, while the diffusion constant is varied from $10^6$ to $10^{10}$\,s$^{-1}$. The blue and orange vertical arrows indicate the values of the correlation time for rotamer jumps, when populations are equal and unequal, respectively.}
	\label{fig:ContributionMotionsAllTauc}
\end{figure}

\begin{figure}[!ht]
	\begin{center}
		\includegraphics[width=0.9\textwidth]{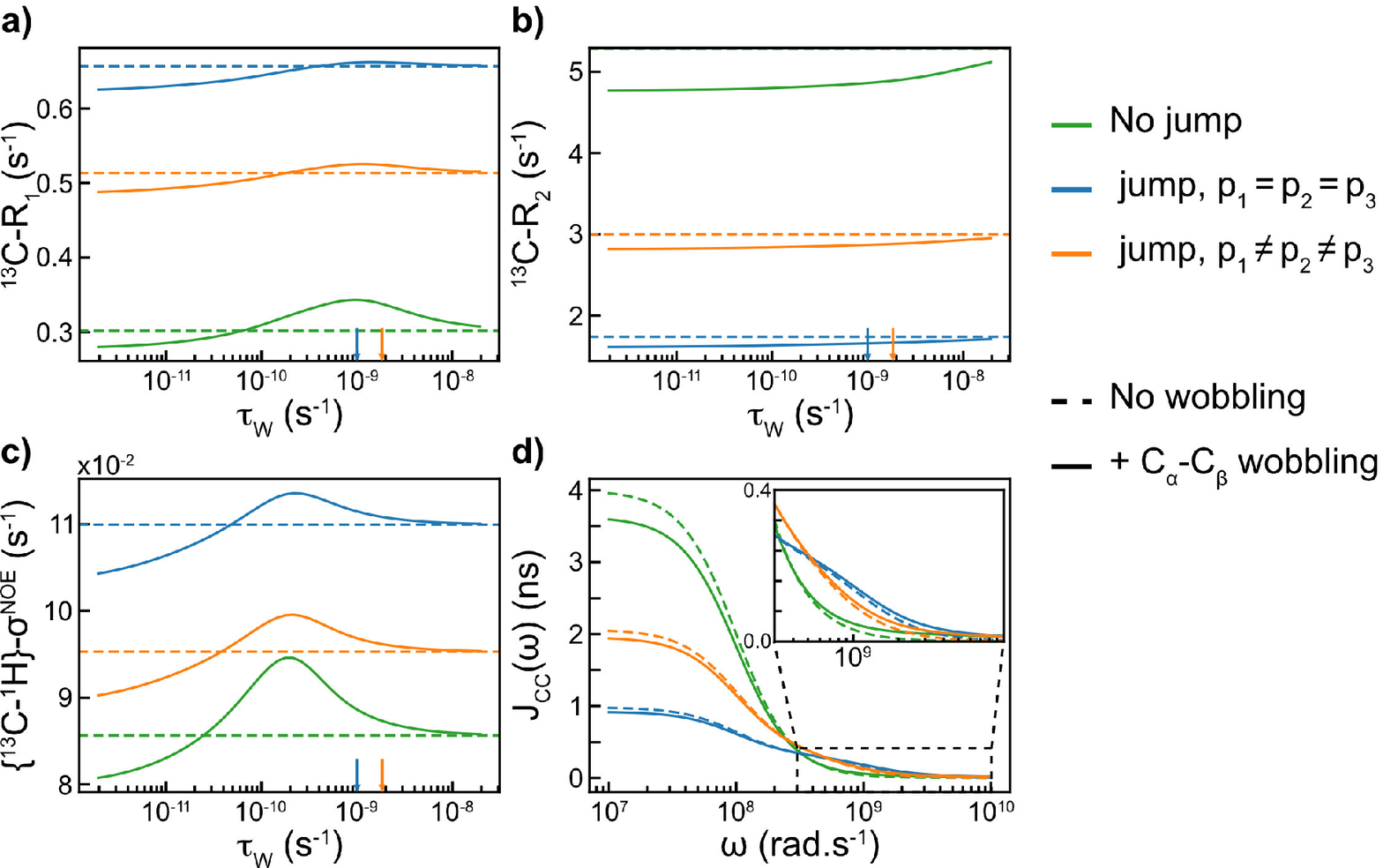}
	\end{center}
	\caption{Contribution of rotamer jump and wobbling in a cone motions to relaxation. Carbon R$_1$ (\textbf{a}), R$_2$ (\textbf{b}) and carbon-proton $\sigma^\mathrm{NOE}$ (\textbf{c}) at 14.1\,T in a hypothetical $^{13}$C$^1$H$^2$H$_2$ valine methyl group, and associated spectral density function for the C$_{\beta}$-C$_{\gamma1}$ bond auto-correlations. The model of motion includes the methyl group rotation as well as (when mentioned) rotamer jump of the C$_\beta$-C$_\gamma$ bond and (when mentioned) the wobbling in a cone of the C$_\alpha$-C$_\beta$ bond. Dashed lines show the value of the relaxation rates and spectral density function without wobbling. Relaxation rates (\textbf{a}-\textbf{c}) and spectral density function (\textbf{d}) calculated in the abscence of rotamer jump are shown in green. In the presence of rotamer jump, the case where all rotamer populations are equal (blue) and unequal (orange) are distinguished. In this latter case, populations are $p_1=0.7$, $p_2=0.2$ and $p_1=0.1$. Calculations are shown as a function of the correlation time for wobbling, a function of the wobbling diffusion constant and cone semi-angle opening, set here to $\beta_{c}=15$\,deg, while the diffusion constant $D_W$ is varied from $10^6$ to $10^{10}$\,s$^{-1}$. The blue and orange vertical arrows indicate the values of the correlation time for rotamer jump, respectively when populations are equal and unequal. The spectral density functions are shown for $D_W = 10^8$\,s$^{-1}$.}
	\label{fig:ContributionMotionsWER}
\end{figure}

\begin{table*}[t]
	\caption{Values of the fitted MF parameters (correlation function Eq.\,\ref{eq:GenMFCorrFunc}) for 4 types of internal motions and the three possible symmetry properties of the overal diffusion tensor. In the case of the symmetric top and fully asymmetric overal diffusion tensor, we report the average and standard deviation over all possible orientations, \textit{i.e.} the different values of $\theta_{D,M}$ and $\{ \varphi_{D,M},\theta_{D,M} \}$ respectively. In these fits, the values and orientation of the diffusion tensor (angle $\Omega_{D,M}$) were fixed such that $\mathcal{S}_{MF}^2$ and $\tau_{MF}$ were the only adjustable parameters. The diffusion tensor for global tumbling has $D_{xx}=10^7$\,s$^{-1}$, $D_{yy}=5\times10^7$\,s$^{-1}$ and $D_{zz}=10^8$\,s$^{-1}$. When the symmetric model is used, eigenvalues are $D_\parallel = D_{zz}$ and $D_\perp = \frac{1}{2} (D_{xx} + D_{yy})$. For the isotropic diffusion, the eigenvalue equals $D=\frac{1}{3}(D_{xx}+D_{yy}+D_{zz})$. The orientation of the rotamer frames in the jump frame are $\Omega_{J,R}^{(2)}=\{\pm\frac{\pi}{2},\beta_{CC}\}$ when two states are considered, and $\Omega_{J,R}^{(3)}=\{2n\frac{\pi}{3},\beta_{CC}\}$ for the 3-state jump with $n=0, 1, 2$ and where $\beta_{CC}=76^\circ$. The 3-state jump model with unequal populations (bottom row) was simulated with $p_1 = 0.7$, $p_2 = 0.2$ and $p_3=0.1$. The rotation on a cone is modeled assuming a methyl group gegometry and with $D_{rot}=10^{11}$\,s$^{-1}$. For the wobbling motion, the cone semi-angle opening is $\beta_c = 20^\circ$ and $D_W = 10^8$\,s$^{-1}$.}
	\begin{center}
		{\def\arraystretch{1.5}
		\begin{tabular}{|c|c|c|c|c|c|c|}
			\hline
		 	 & \multicolumn{2}{c|}{Isotropic} & \multicolumn{2}{c|}{Symmetric top} & \multicolumn{2}{c|}{Fully asymmetric} \\%
			\hline
			& $\mathcal{S}_{MF}^2$ & $\tau_{MF}$ (ps) & $\mathcal{S}_{MF}^2$  & $\tau_{MF}$ (ps)& $\mathcal{S}_{MF}^2$ & $\tau_{MF}$ (ps)\\ 
			\hline
			rotation & $0.12$ & $4.5$ & $0.12 \pm 0.00$ & $4.7 \pm 0.0$ & $0.11 \pm 0.00$ & $4.9 \pm 0.0$ \\
			wobbling & $0.83$ & $340$ &$0.83 \pm 0.00$ & $348 \pm 27$ & $0.83 \pm 0.00$ & $351 \pm 27$ \\
			2-state jump &  \multirow{2}{*}{$0.82$} &  \multirow{2}{*}{$612$} &  \multirow{2}{*}{$0.56 \pm 0.24$} &  \multirow{2}{*}{$1,368 \pm 709$} &  \multirow{2}{*}{$0.52 \pm 0.29$} &  \multirow{2}{*}{$1,840 \pm 1, 181$} \\
			$p_1=p_2$ & & & & & & \\
			3-state jump & \multirow{2}{*}{$0.16$} & \multirow{2}{*}{$319$} & \multirow{2}{*}{$0.16 \pm 0.01$} & \multirow{2}{*}{$331 \pm 21$} & \multirow{2}{*}{$0.17 \pm 0.00$} & \multirow{2}{*}{$329 \pm 20$} \\
			$p_1=p_2=p_3$ & & & & & & \\
			\hline
			\hline
			3-state jump & \multirow{2}{*}{$0.43$} & \multirow{2}{*}{$153$} & \multirow{2}{*}{$0.43 \pm 0.11$} & \multirow{2}{*}{$188 \pm 117$} & \multirow{2}{*}{$0.42 \pm 0.12$} & \multirow{2}{*}{$186 \pm 169$} \\
			$p_1 \neq p_2 \neq p_3$ & & & & & & \\
			\hline
		\end{tabular}
		}
	\end{center}
	\label{table:FitParamMF_FakeDiffTensor}
\end{table*}